\documentclass[10pt]{article}
\usepackage{palatino}
\usepackage{epsfig}
\usepackage{amsmath}
\def\sec#1{\vspace{\baselineskip}\noindent\begin{large}%
{\bf{\textsf{#1}}}\end{large}\nopagebreak[3]\vspace{\medskipamount}}
\def\subsec#1{\vspace{\baselineskip}\noindent\begin{large}%
{\textsf{#1}}\end{large}\nopagebreak[3]\vspace{\smallskipamount}}

\def\b{\begin{equation}}
\def\e{\end{equation}}
\def\ba{\begin{eqnarray}}
\def\ea{\end{eqnarray}}
\def\mtx#1{\mathbf{#1}}
\def\avg#1{\left<#1\right>}
\def\ol#1{\overline{#1}}
\def\tsp{^{\mathrm{T}}}
\def\eq{\!=\!}
\def\deg{$^{\circ}$}
\def\rfs#1{ [\ref{#1}]}
\def\rfd#1#2{ [\ref{#1},\ref{#2}]}

\def\rfr#1#2{ [\ref{#1}--\ref{#2}]}
\def\rfrr#1#2#3#4{ [\ref{#1}--\ref{#2},\ref{#3}--\ref{#4}]}

\def\rfsrd#1#2#3#4#5{ [\ref{#1},\ref{#2}--\ref{#3},\ref{#4},\ref{#5}]}
\def\EL{$E_L$}
\def\EB{$E_B$}
\def\FI{$\mtx{\Phi}$}
\def\FF{$\mtx{F}$}
\def\Fvec{$\vec{F}_{\alpha}$}
\def\MM{$M \! \times \! M$}
\def\nn{$n \! \times \! n$}
\def\nM{$n \! \times \! M$}

\newcounter{bean}
\newenvironment{refs}%
  {\begin{list}{\hfill \arabic{bean}.}{%
    \usecounter{bean}
    \setlength{\leftmargin}{0.2in}%
    \addtolength{\labelsep}{0.00in}
    \addtolength{\leftmargin}{\labelsep}
    \setlength{\itemindent}{0in}%
    \setlength{\labelwidth}{2em}%
    \setlength{\rightmargin}{0.0in}%
    \setlength{\parsep}{0.0in}%
    \setlength{\itemsep}{5pt}%
    \setlength{\parindent}{0.0in}%
    \setlength{\partopsep}{0.0in}%
    \setlength{\topsep}{0.0in}}}%
  {\end{list}} 

\begin{document}

\begin{sf}

\begin{Large}
\noindent
\textbf{How Behavioral Constraints May Determine Optimal 
Sensory Representations \\ }
\end{Large}

\vspace{\smallskipamount} 

\noindent
\textbf{Emilio Salinas} \\

\begin{small}
\noindent
Department of Neurobiology and Anatomy, 
Wake Forest University School of Medicine,
Winston-Salem, NC 27157-1010, USA\@.
\textbf{E-mail:} esalinas@wfubmc.edu \\
\end{small}

\vspace{\smallskipamount} 

\noindent
The sensory-triggered activity of a neuron is typically
characterized in terms of a tuning curve, which describes the neuron's
average response as a function of a parameter that characterizes a
physical stimulus. What determines the shapes of tuning curves in a
neuronal population? Previous theoretical studies and related
experiments suggest that many response characteristics of sensory
neurons are optimal for encoding stimulus-related information.  This
notion, however, does not explain the two general types of tuning
profiles that are commonly observed: unimodal and monotonic. Here, I
quantify the efficacy of a set of tuning curves according to the
possible downstream motor responses that can be constructed from them.
Curves that are optimal in this sense may have monotonic or
non-monotonic profiles, where the proportion of monotonic curves and
the optimal tuning curve width depend on the general properties of the
target downstream functions. This dependence explains intriguing
features of visual cells that are sensitive to binocular disparity and
of neurons tuned to echo delay in bats. The numerical results suggest
that optimal sensory tuning curves are shaped not only by stimulus
statistics and signal-to-noise properties, but also according to their
impact on downstream neural circuits and, ultimately, on behavior. \\

\vspace{\smallskipamount} 

\begin{small}
\noindent
Citation: Salinas E (2006) How behavioral constraints may determine
optimal sensory representations. PLoS Biol 4(12): e387\@. 
DOI: 10.1371/journal.pbio.0040387 \\
\end{small}

\end{sf}


\sec{Introduction}

Sensory neurons respond to physical stimuli, and this relationship is
often quantified by plotting their evoked activity --- for instance,
the mean firing rate --- as a function of a relevant stimulus
parameter. The resulting response functions or tuning curves have been
the subject of much theoretical work, particularly in vision. In
trying to understand such tuning curves, the emphasis has been on
information maximization, the main idea being that sensory neurons
should represent the sensory world as accurately and efficiently as
possible\rfr{Atic92}{Simo03}. This principled approach, known as the
efficient coding hypothesis, has been extremely successful at
predicting the receptive field properties of neurons in early
visual\rfr{AR90}{BS97} and auditory\rfd{Lewi02}{SL06} areas, and is
consistent with numerous experimental observations\rfr{DAR96}{CWT04}. 

However, information maximization is not enough. Such a principle
cannot completely account for the response characteristics of cortical
neurons, particularly beyond early sensory areas, because it does not
consider how the encoded information will be used, if at all --- it
would not make sense for sensory neurons to pack lots of information
into parts of feature space that are of little relevance to the
animal. A recent study\rfs{MGLH05} investigating auditory responses in
grasshoppers illustrates this.  Primary auditory receptors in
grasshoppers do not respond equally well to different kinds of
environmental sounds. Instead, the stimulus ensemble that maximizes
their information rate consists of short segments of grasshopper songs
that mark the transitions between song syllables\rfs{MGLH05}. Thus,
such early receptor neurons seem to be highly specialized for
describing a rather small set of sounds that are relevant for a
specific behavior, namely, discriminating grasshopper
songs\rfs{MSFKSRH03}. 

This raises an interesting question, does an animal's behavior
influence the shapes of its sensory tuning curves? and if so, what
features would be most sensitive to behavioral constraints? There are,
in fact, two motivations for addressing this problem. On one hand, the
limitations just discussed of the efficient coding principle. On the
other, what I see as a theoretical mystery: the ubiquity of monotonic
tuning curves. Tuning curves come in two main flavors, single-peaked
and monotonic (increasing or decreasing). Bell-shaped curves with a
single peak are the textbook example of tuning functions. They are
indeed quite common\rfr{Albr84}{O'KB96}, and many modeling studies
have investigated the coding properties of arrays of such unimodal
curves subject to some form of noise\rfr{Para88}{BG06}. Monotonic
dependencies on stimulus parameters, however, have also been amply
documented, not only in the somatosensory system\rfr{RBHL99}{SHZR00}
but also in other modalities\rfr{BITDH97}{KBOdBV05}. Monotonic tuning
curves have received little attention from theorists. No analysis has
been reported from the standpoint of efficient coding, and it is not
clear whether they present any advantage regarding other criteria,
such as learning\rfs{Guig03}. To complicate matters further, some
neuronal populations show mixtures of monotonic and peaked
curves\rfr{HC05}{PvE05}.

Why is there such a range of tuning curve shapes? And, in particular,
what promotes the development of monotonic profiles?  To investigate
more closely whether behavioral factors play a role in this problem,
here I evaluate the responses of a neuronal population not only in
relation to their sensory inputs but also in terms of the range of
outputs that they are capable of generating. The sensory tuning curves
are seen as a set of basis functions from which other functions of the
stimulus parameters can be easily constructed\rfd{Pogg90}{PS97}; these
other functions represent motor activity or actions that are generated
in response to a stimulus. The idea is that, if something can be said
about the statistics of the downstream motor activity, then we should
be able to say something about the sensory tuning curves that are
optimal for driving such activity. 


\sec{Results}

\vspace{-\baselineskip}
\subsec{Tuning Curves as Basis Functions}

To begin, the problem needs to be defined mathematically. The
situation can be described using some of the tools of classic function
approximation\rfd{PG89}{SA00}, and is schematized in Figure~1: $n$
basis neurons respond to $M$ stimuli or conditions and drive $N$
additional downstream neurons whose output should approximate a set of
desired functions \FF. 
\begin{figure*}[t!]
\centerline{\epsfig{figure=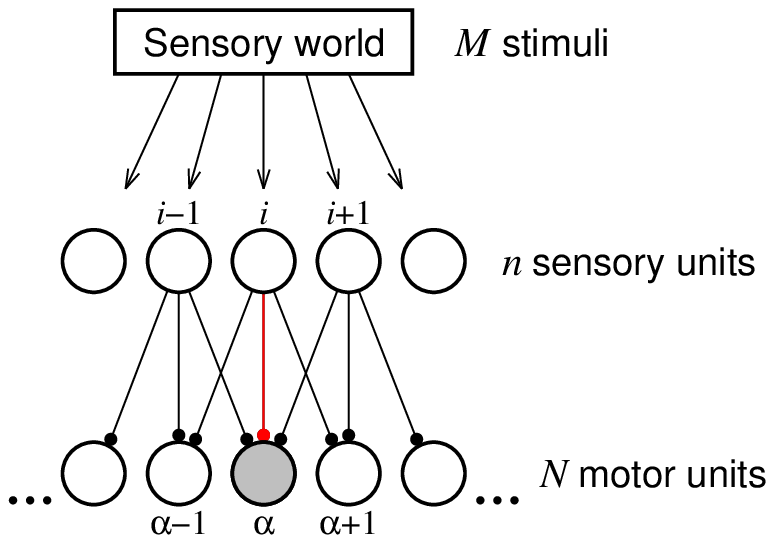,width=3.1in,clip} \hspace{0.10in}
\parbox[b]{2.80in}{\small
{\sf \textbf{Figure 1\@.}} 
Schematic of the model.
There are $n$ sensory or basis neurons that respond to $M$ stimuli and
drive $N$ motor neurons downstream. The firing rate of motor neuron
$\alpha$ (shown filled) when stimulus $k$ is presented is equal to 
$R_{\alpha k} = \sum_{i=1}^n w_{\alpha i} \, r_{ik}$,
where $r_{ik}$ is the firing rate of sensory neuron $i$ and 
$w_{\alpha i}$ is the connection (shown in red) from sensory neuron
$i$ to downstream neuron $\alpha$. For each motor neuron $\alpha$, the
driven response $R_{\alpha k}$ should approximate as closely as
possible a desired response $F_{\alpha k}$. \vspace{4pt}}}
\end{figure*}
The basis neurons represent sensory neurons
whose tuning curves we are interested in, and the downstream units
represent motor neurons that contribute to generating actions. The key
quantity to study is the matrix $\mtx{r}$, where $r_{ik}$ is the
firing rate of basis neuron $i$ evoked by stimulus $k$.  These basis
responses may have intrinsic variability (noise), so their mean values
are denoted as $\avg{r_{ik}}$, where the brackets indicate an average
over multiple presentations of the same stimulus. Since the second
index parameterizes stimulus values, the tuning curve of cell $i$ is
simply $\avg{r_{ik}}$ plotted as a function of $k$. As mentioned
above, the rationale of this approach is that, although the motor
responses \FF\ may be largely unknown in reality, if they have some
regularity or statistical structure, this should partially determine
the optimal shapes of the sensory tuning curves $\avg{\mtx{r}}$. For
the moment, however, pretend that the repertoire of motor responses
\FF\ that should be elicited by the stimuli is fully known.

To proceed, a mechanism is needed for the sensory neurons to
communicate with the motor neurons. The simplest assumption is that
the downstream motor units are driven through weighted sums. Thus, 
the response of downstream unit $\alpha$ to stimulus $k$ is
$R_{\alpha k} \eq \sum_{i=1}^n w_{\alpha i} \, r_{ik}$, 
where $w_{\alpha i}$ represents the synaptic connection from sensory
neuron $i$ to downstream neuron $\alpha$ (Figure~1). In matrix
notation, this is
$\mtx{R} \eq \mtx{w} \, \mtx{r}$.
In this simple model, the shapes of the tuning curves become important
when there are more downstream neurons than basis neurons ($n<N$) and
when there is noise, so both conditions are assumed to be true.  

Next, recall that the job of downstream unit $\alpha$ is to produce
the target motor response \Fvec\ (where \Fvec\ is row $\alpha$ of
\FF). Therefore, what is needed is for the driven responses,
$\mtx{R} \eq \mtx{w} \, \mtx{r}$,
to approximate as closely as possible the desired ones, \FF.
Crucially, however, different sets of tuning curves $\avg{\mtx{r}}$
will vary in their capacity to generate the target downstream
responses. This capacity is quantified using an error measure denoted
as \EB. When \EB\ is zero, the sensory (basis) neurons are most
accurate and the driven responses are equal to the desired ones; when
\EB\ is 1, the driven activity has little or no resemblance to the
desired activity and the error is maximal.  The derivation of \EB\ is
presented in the Methods section. What is important, however, is to
understand its dependencies, which are as follows: 
$E_B = E_B(\avg{\mtx{r}}, \boldsymbol{\sigma}, \{s_k\}, \mtx{\Phi})$.
First, the error depends on the sensory tuning curves
$\avg{\mtx{r}}$ and on their noise, $\boldsymbol{\sigma}$. Second,
note that there is no dependence on the synaptic weights. This is
because \EB\ is constructed assuming that, for each $\avg{\mtx{r}}$, 
the best possible synaptic weights are always used. Third, \EB\
depends on how often each stimulus is shown; that is, on the set
of coefficients $\{s_k\}$, where $s_k$ is the probability that
stimulus $k$ is presented.  Finally, \EB\ does not depend directly on
the actual motor responses $\mtx{F}$. Instead, the key independent
quantity is their correlation matrix \FI, which captures their overall
statistical structure. Its components are
\b
   \Phi_{kl} = \frac{1}{N} \, 
               \sum_{\alpha=1}^N F_{\alpha k} \, F_{\alpha l} . 
\e
In essence, \FI\ represents an average over all the downstream motor
responses that the basis neurons have to approximate. This average
corresponds to drawing the $F_{\alpha k}$ values from given
distributions, or equivalently, to choosing multiple functions \Fvec\
from a given class (see below). 

In summary, given the noise of the neurons ($\boldsymbol{\sigma}$),
the statistics of the stimuli ($\{s_k\}$), and the statistics of the
downstream responses ($\mtx{\Phi}$), the error \EB\ can be calculated
for any set of sensory tuning curves $\avg{\mtx{r}}$.

\subsec{What Determines the Optimal Tuning Curves?}

So far, what I have done is set up the problem and develop a quantity
that measures the effectiveness of the sensory tuning curves as
building blocks for constructing the desired motor responses.  Recall,
however, that the goal is to find the best tuning curves. In the
present formalism, this is the same as asking what tuning curves
$\avg{\mtx{r}}$ minimize \EB. 

It should be noted, however, that \EB\ cannot completely determine the
optimal tuning curves. This is because the problem is fundamentally
under-constrained: since the network model is linear 
($\mtx{R} \eq \mtx{w} \, \mtx{r}$),
any transformation by an invertible matrix $\mtx{A}$ such that 
$\mtx{w} \rightarrow \mtx{w} \mtx{A}$ 
and
$\mtx{r} \rightarrow \mtx{A}^{-1} \mtx{r}$ 
produces the same approximation and thus leaves the error unchanged.
Therefore, additional conditions on $\mtx{w}$ or $\mtx{r}$ are
required to make the solution unique. These conditions are crucial, in
that they can lead to quite different results\rfd{OF96}{LS99}, but it
is instructive to ignore them momentarily; this provides some
intuition into the problem, as well as a lower bound on \EB.

Before considering specific examples, it is important to discuss the
key factors that will determine the solution. Intuitively, the tuning
curves should match, as much as possible, the $N$ target functions
\Fvec. If all the functions are different, then clearly a lot of
tuning curves will be needed for accurate approximation. In this case,
`different' means `not highly correlated', which in turn means that
\FI\ will have large values along its diagonal (see below). On the
other hand, if the functions \Fvec\ are similar to each other, then
very few tuning curves should suffice. Or, if more tuning curves are
available, many of them can be used to cover specific regions where
\FI\ varies more abruptly. Therefore, what matters when designing
tuning curves is really the number of distinct functions that need to
be approximated, as measured both by how big $N$ is and how correlated
the functions \Fvec\ are.

The rest of this section formalizes this intuition and describes more
precisely the dependence of the optimal tuning curves on \FI. The
reader who wishes to skip the mathematical details may safely
move on to the next section.

To better understand the effect of \FI, it is useful to decompose it
using a special set of vectors (eigenvectors) and their corresponding
coefficients (eigenvalues). The idea is to use the eigenvectors of
\FI\ to construct the optimal tuning curves. Assuming that all stimuli
are equally probable, the key property of \FI\ is that its $M$
eigenvalues are all non-negative and add up to $M$ (see Methods).
When \FI\ results from averaging either just a few functions 
($\ll \! M$) or many functions with similar shapes, only a few
eigenvalues are significantly larger than zero. Conversely, when the
average involves many different functions, most eigenvalues are close
to 1 and \FI\ is strongly diagonal. 

Keeping these properties in mind, as well as the fact that \EB\ varies
between 0 and 1, now consider a single basis neuron.  Assume that its
tuning curve is proportional to an eigenvector of \FI\ with eigenvalue
$\lambda$. In that case, \EB\ depends on only two numbers, $\lambda$
and a signal-to-noise ratio $\rho$ that is equal to the mean
response squared divided by the mean variance of the neuron. That is,
\b
    E_B(n\eq 1) = \frac{ \left(1 - \frac{\lambda}{M}\right) 
                           \rho + 1 }
                         { \rho + 1 } 
    \label{EBerr1}
\e
(see Methods).  This expression leads to three important observations. 
(1) When the neuron's variance increases, $\rho$ tends to zero and the
error tends to 1. Thus, as expected, higher noise always pushes the
error toward its maximum. 
(2) The worst-case scenario is $\lambda \eq 0$. This produces the
maximum error, regardless of the noise, and occurs when the tuning
curve is completely different from (orthogonal to) all the target
functions used to compute \FI.
(3) For any signal-to-noise ratio, the lowest error occurs when
$\lambda$ is the largest eigenvalue of \FI, in which case the single
tuning curve is equal to the so-called first principal
component\rfs{Joll02} of \FI. This one tuning curve may suffice to
generate a very small error, if the noise is low and 
$\lambda \eq \lambda^{\mathrm{max}} \approx M$. 
But, on the other hand, if $\lambda^{\mathrm{max}}$ is small, the
error will be large even if the tuning curve has the optimal shape and
zero noise.

The efficacy of the single basis neuron thus depends on its
variability, on the largest eigenvalue of \FI, and on the similarity
between the tuning curve and the eigenvectors. An analogous result is
obtained with more neurons, except that additional eigenvalues and
eigenvectors become involved (see Methods).  Specifically, with $n$
basis neurons and no noise, the minimum error that can be achieved is 
\b
    \mbox{min} \! \left( E_B \right) = 1 - \frac{1}{M} \, 
                                           \sum_{i=1}^{n} \lambda_i 
    \label{EBerr2}
\e
where $\lambda_1,\ldots \lambda_{n}$ are the $n$ largest eigenvalues
of \FI. The key in this expression is that the sum involves $n$ terms
only.  This is significant because, if \FI\ has just a few large
eigenvalues, the sum of the $n$ largest ones may approach $M$ even if
$n \ll \! M$, so few noiseless tuning curves with the right shapes
will suffice for representing accurately all the desired motor
responses. This happens, for instance, when the motor responses are
similar to each other, i.e., are highly correlated. Conversely, if
many eigenvalues are close to 1, then \FI\ is strongly diagonal and it
is certain that a much larger number of sensory neurons will be
needed, even if noise is not a factor.  Numerical results support
these theoretical conclusions (see Supporting Information). 

\subsec{Monotonic Versus Non-Monotonic Tuning Curves}

Armed with a criterion that quantifies the accuracy of the sensory
tuning curves and takes into account the statistics of the motor
outputs, now we can ask, what sets of tuning curves are optimal when
there is variability and specific families of downstream functions are
considered? To investigate this, each tuning curve was parameterized
by four numbers, such that either monotonic or unimodal profiles with
a large variety of shapes could be produced, and a numerical routine
was used to find optimal parameter combinations that minimized \EB\
(see Methods). By limiting the possible tuning curve shapes, this
procedure eliminated the ambiguity problem mentioned earlier. 

\begin{figure*}[t!]
\centerline{\epsfig{figure=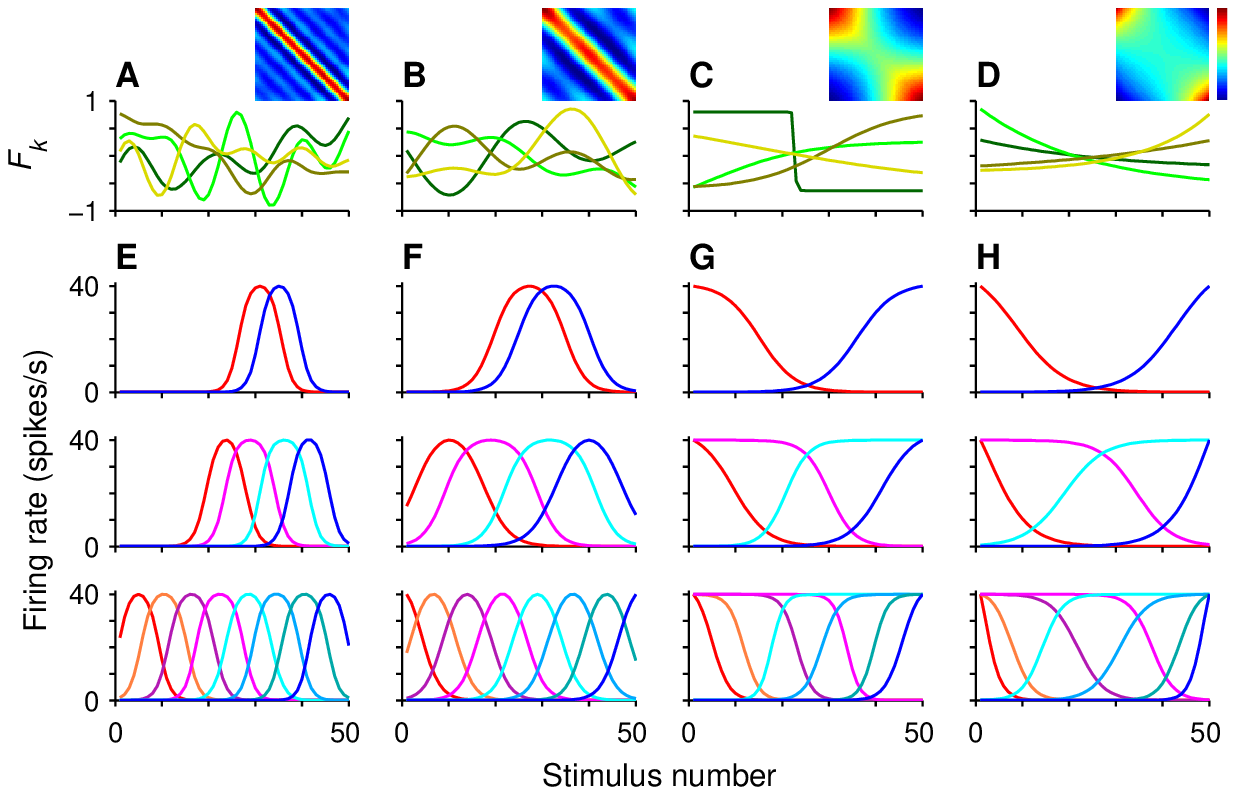,width=4.3in,clip}}
\vspace{0.1in}
\parbox[t]{\textwidth}{\small
{\sf \textbf{Figure 2\@.}}
Optimal tuning curves for four classes of downstream functions.
\textsf{(A)} High-frequency oscillating functions. Each function
$\vec{F}$ was composed of 8 sinusoids of random phase and amplitude.  
Four examples are shown. Inset depicts correlation matrix obtained
from 5000 functions.
\textsf{(B)} Low-frequency oscillating functions. 
\textsf{(C)} Saturating monotonic functions. Each $\vec{F}$ was an
increasing or decreasing sigmoidal curve of random steepness and
center point.
\textsf{(D)} Non-saturating monotonic functions. Each $\vec{F}$ was an
increasing or decreasing exponential curve with random steepness.
\textsf{(E}--\textsf{H)} Optimal sets of 2, 4 and 8 tuning curves for
the classes in the corresponding columns. Shown responses minimized
\EB\ and were constrained to remain between 0 and 40 spikes/s. }
\end{figure*}

Figure~2 illustrates the results of computer experiments in which
optimal tuning curves were obtained numerically for four classes of
downstream responses. Examples of functions within each class are
shown on the top row, next to the corresponding \FI\ matrices
(Figure~2A--D). The graphs below show the sets of 2, 4 and 8 tuning
curves that minimized \EB\ in each case.  When the target functions
are non-monotonic (Figure~2A and 2B), the optimal tuning curves are
themselves non-monotonic (Figure~2E and 2F). Similarly, when the
target functions are monotonic (Figure~2C and 2D), the optimal tuning
curves are also monotonic (Figure~2G and 2H), even though the target
classes comprise both increasing and decreasing functions. The
detailed features of the optimal tuning curves clearly depend on the
specifics of the target class.  For instance, the number of peaks and
troughs of the oscillating target functions affects the optimal width
of the unimodal curves (compare Figure~2E and Figure~2F).  Most
notably, however, because the noise properties and stimulus statistics
remained constant, all the differences across columns are due to
constraints that act downstream from the sensory neurons. 

These results were highly robust with respect to various manipulations
(see Supporting Information). Increasing the noise, adding a power
constraint, using non-uniform stimulus probabilities, or
parameterizing the tuning curves differently did not alter the main
finding: optimal tuning curves are predominantly monotonic or
non-monotonic depending on the type of downstream activity they are
meant to evoke. Furthermore, manipulating the stimulus probabilities
alone never gave rise to monotonic curves; for this, a monotonic trend
in the downstream responses was necessary.

As mentioned in the Introduction, both unimodal and monotonic tuning
curves are found in various parts of the brain, and this diversity has
remained unexplained (see also the Discussion). The above results
suggest that the two types of responses may arise not because of
information-coding considerations but because of differences in the
actions that various types of stimuli ultimately trigger. For
instance, some stimulus parameters, such as the orientation of a bar,
should lead to approximately the same sorts of movements regardless of
the parameter's value. But other parameters or features, such as image
contrast or sound intensity, have an obvious directionality, in that
salient stimuli of high contrast or high intensity are more likely to
lead to action. Thus, sensory neurons might respond in a qualitatively
different way to features with and without such a behavioral bias
because that is the most effective way to generate the appropriate
actions. The next two sections present two realistic situations where
such motor asymmetries may arise.

\subsec{Mixed Tuning Curves for Binocular Disparity}

Binocular disparity provides an interesting example of a signal that
is likely associated with an intrinsic bias in behavior. To see the
source of the asymmetry, consider what possible movements may be
triggered by a visual stimulus at a given disparity. If a stimulus is
seen near zero disparity (i.e., at the plane of fixation), many
subsequent actions are possible, such as reaching, biting, fixating,
etc. In contrast, if a relevant stimulus appears at a positive
disparity (i.e., behind the plane of fixation), a diverging eye
movement should typically follow, because that will bring the object
onto the plane of fixation for more detailed examination. Conversely,
converging eye movements should be seen more often following stimuli
of negative disparity (i.e., in front of the plane of fixation). As a
consequence, an oculomotor unit that is strongly activated at positive
disparities should have a tendency to fire weakly at negative
disparities, and viceversa. This rationale implies that, for any
relevant oculomotor cell, the responses at opposite ends of the
disparity range should tend to be anti-correlated, and these should be
approximately independent of the responses triggered near zero
disparity. The downstream functions in Figure~3A are meant to capture
this statistical regularity. They vary strongly in the middle of the
stimulus range but have much more stereotyped values at the extremes.

\begin{figure*}[t!]
\centerline{\epsfig{figure=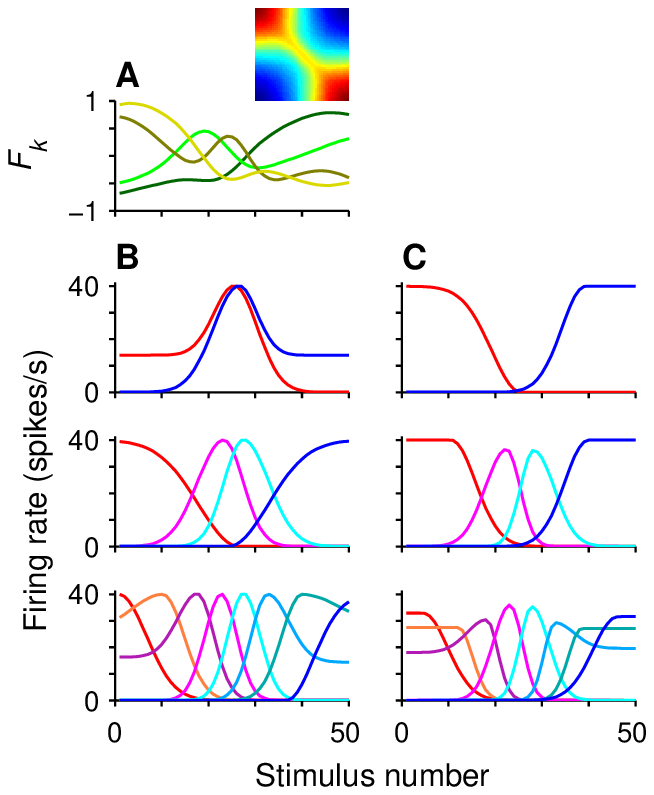,width=2.8in,clip} \hspace{0.15in}
\parbox[b]{3.05in}{\small
{\sf \textbf{Figure 3\@.}} 
Optimal tuning curves for downstream functions that have both
peaked and monotonic components. 
\textsf{(A)} Four examples of functions $\vec{F}$ obtained by
combining a localized oscillatory function (with a Gaussian envelope)
and a saturating monotonic function. Such functions represent
hypothetical motor responses to stimuli at various binocular
disparities. Inset depicts correlation matrix.
\textsf{(B)} Optimal sets of 2, 4 and 8 sensory tuning curves obtained
with low noise.
\textsf{(C)} As in \textsf{(B)}, but with high noise and high power
cost.  In all plots, the x-axis represents binocular disparity.
\vspace{4pt} }}  
\end{figure*}

Optimal tuning curves for this class of downstream functions are shown
in Figure~3B and 3C. These curves have two novel features: they mix
unimodal and monotonic profiles, and include intermediate curves with
a peak superimposed on a monotonic component.  Recently, it has been
shown that disparity tuning curves in area V4 have precisely these
characteristics. The V4 population comprises a continuum of disparity
tuning patterns that includes monotonic (the classic near and far
cells), unimodal (the classic tuned cells) and intermediate
cells\rfs{HC05}. 

\subsec{Widening Tuning Curves for Echo Delay}

The final example addresses the issue of tuning curve width. The
downstream functions illustrated in Figure~4A are meant to capture a
distinctive aspect of the behavior of bats, which locate prey by means
of echolocation. In this case, consider a bat pursuing a moth.  From
far away, the bat can approach the moth by following its average path,
smoothing out the moth's high-frequency maneuvers.  At a close
distance, however, the bat must turn at least as sharply as the moth
itself in order to catch it, particularly in a cluttered
environment\rfd{GM06}{MBGS06}. Thus --- this is the crucial assumption
--- when a bat flies toward a small target, its maneuvers must be
faster as the target is approached.  This postulate is translated into
a statement about motor responses by generating functions that vary
rapidly near stimulus 1 (corresponding to near targets, or short echo
delays) and vary progressively more slowly at higher stimuli
(corresponding to far targets, or long echo delays).  Examples of such
hypothetical motor responses are shown in Figure~4A.  The optimal
tuning curves for this case are non-monotonic, as might have been
expected, but most notably, their widths increase as functions of the
preferred stimuli (Figure~4B).  This effect is extremely robust. It
was also observed when the tuning curves were parameterized
differently, and when high noise and high power cost were used
(Figure~4C). Many auditory neurons of the bat's sonar system have this
particular property. They are tuned to echo delay, and their
tuning-curve widths vary linearly with the so-called `best
delay'\rfd{SH86}{OS91}, which is the echo delay at which the peak
response is elicited. 

\begin{figure*}[t!]
\centerline{\epsfig{figure=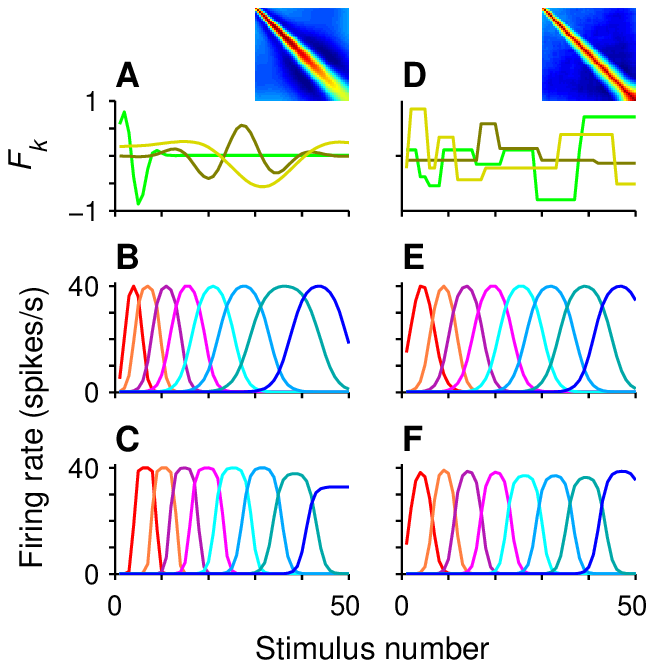,width=2.8in,clip} \hspace{0.15in}
\parbox[b]{3.05in}{\small
{\sf \textbf{Figure 4\@.}} 
Optimal tuning curves for downstream functions that vary more
rapidly near one end of the stimulus range.
\textsf{(A)} Three examples of continuous, oscillatory functions (with
Gaussian envelopes) that oscillate at high frequency near stimulus 1
and at progressively lower frequency near stimulus 50. They represent
hypothetical motor responses of bats as functions of echo delay or
target distance.
\textsf{(B)} Set of 8 tuning curves that minimized \EB\ given the
correlation matrix in \textsf{(A)} and low noise. 
\textsf{(C)} As in \textsf{(B)}, but with high noise and high power
cost.
\textsf{(D)} Three examples of discontinuous $\vec{F}$ functions. Each
one is a collection of constant segments placed randomly. Segment
width increased linearly as a function of segment location on the
x-axis.
\textsf{(E)} Set of 8 tuning curves that minimized \EB\ given the
correlation matrix in \textsf{(D)} and low noise. 
\textsf{(F)} As in \textsf{(E)}, but with high noise and high power
cost. In all plots, the x-axis represents echo delay.
\vspace{3pt} }}  
\end{figure*}

Again, note that the model generates this result based on a single
statistical assumption about the motor responses, which is a
progressive change in their absolute rate of variation along the
echo-delay range. This is confirmed by Figure~4D, for which radically
different downstream functions were generated. Here,
piecewise-constant functions were used, each composed of a variable
number of segments that had random amplitudes and locations. The only
structure was a correlation between segment length and segment
location along the x-axis. It is this correlation that gives rise to
the systematic change in tuning curve width (Figure~4E and 4F).

A key question here, however, is whether curves of increasing width
could also result from an uneven distribution of stimulus
probabilities $s_k$, without assuming an asymmetry in the downstream
functions. Technically the answer is yes --- a progressive widening
was obtained by using monotonically increasing stimulus probabilities
together with the downstream functions in Figure~2A and 2B. But there
were three severe problems with this purely sensory mechanism: (1) the
effect required high noise, (2) it was much weaker, meaning that
variations in width were small, and (3) most importantly, it placed
the narrow tuning curves in the region of highly probable stimuli,
which for the bat means that nearby targets must be encountered much
more often than far away ones.  Therefore, the puzzling widening of
sensory tuning curves documented in the bat may be explained more
parsimoniously by assuming that flight control needs to be faster as
the target gets closer.

\subsec{Other Tuning Curve Shapes}

The parametric approach presented here allows a direct comparison
between monotonic and peaked tuning curves.  Would the results hold,
however, if other shapes were allowed?  To address this question,
optimal tuning curves were recalculated using drastically different
constraints. The basis responses were simply required to be positive
and bounded, whereas the synaptic weights were constrained to be
sparse.  With sparse connectivity, each downstream function is
approximated using only a subset of all the available basis neurons.
The optimal tuning curves obtained with this method were much more
variable, as expected given the absence of restrictions on their
shapes, and often had multiple peaks.  However, an index measuring the
monotonicity of the curves in each population was computed, and in
terms of this index the results were very similar to those obtained
with parameterized curves: the monotonicity of the basis responses was
determined by the monotonicity of the downstream functions, and
conversely, strongly monotonic tuning curves could not be produced by
manipulating the stimulus statistics alone. Details of these numerical
experiments are discussed in the Supporting Information.

\sec{Discussion}

Both unimodal- and monotonic-encoding populations of neurons are
common and are maintained by different brain
regions\rfrr{Albr84}{O'KB96}{RBHL99}{KBOdBV05}, including areas beyond
the periphery where tuning curves seem to be actively
synthesized\rfd{SHZR00}{Adol93}. Yet the factors that determine
whether a specific neuronal population develops monotonic, unimodal,
or mixed responses have remained a mystery. Computationally, unimodal
curves are different from monotonic ones in two ways. First, they
allow learning to be local, in the sense that changing the weight of a
peaked curve affects the output function only over the range of the
curve, not over the entire input space\rfd{Pogg90}{PG89}. Second, it
seems that representing multiple values simultaneously would be much
easier with peaked curves, especially when the difference between
coded values is large relative to the curve width. Although the
importance of these differences remains unclear, they further
illustrate the lack of theoretical justification for monotonic sensory
responses.  A possible solution to this enigma, however, is to
consider the types of actions that various stimuli ultimately trigger.

\subsec{Maximizing Fisher Information Is Not Enough}

The classical approach to sensory coding involves information
maximization\rfs{BG06}. Thus, it would seem that some of the examples
discussed above could be formulated in more familiar terms by
requiring that more Fisher information\rfd{ZS99}{DLP99}, or
equivalently, higher accuracy, be found in certain parts of the
sensory space. For instance, in analogy with Figure~3A, what happens
if much higher accuracy is needed in the middle of the stimulus range
than at the edges? Could such conditions lead to monotonic tuning
curves? 

The answer is no.  This is because a function specifying a desired
relative accuracy at each point in the stimulus range is exactly
equivalent to the set of coefficients $s_k$ that were used to
represent stimulus probabilities. That is, $s_k$ can also be
interpreted as the weight or importance of the error between driven
and desired motor responses when stimulus $k$ is presented (see
Equation~\ref{ELerr}). For instance, when these coefficients had a
Gaussian instead of a uniform profile, the results were entirely
consistent with an increase in Fisher information at the middle of the
range; more tuning curves were located near the middle, and those were
narrower than the ones at the edges. These effects depended on the
level of noise, as expected, and were rather subtle, but the key point
is that such manipulations had no bearing on whether the optimal
tuning curves were monotonic or not (see the Supporting Information
for further results).  Therefore, while information maximization is
clearly important, the downstream functions in this model have a much
stronger influence on the optimal tuning curve shapes.

\subsec{Inputs, Outputs and Optimality}

Previous theoretical studies have attempted to explain the properties
of sensory neurons based on two elements, an optimality assumption
(efficient coding) and the statistics of their inputs, i.e., the
statistics of natural images or natural
sounds\rfsrd{Simo03}{AR92}{SL06}{CWT04}{MGLH05}. Conceptually, the
approach here was not dissimilar. An optimality assumption, accurate
function approximation, together with the statistics of motor
responses were used to infer the shapes of sensory tuning curves.
However, the present model works backwards, in that it requires
knowledge about downstream rather than upstream events (note, however,
that stimulus statistics are still taken into account through the
coefficients $s_k$ and through correlations with the downstream
functions they evoke). Clearly, whereas measuring the statistics of
natural images or sounds is straightforward, determining the
statistics of motor activity associated with specific stimuli poses a
challenge. However, assuming that such motor statistics have some
structure, because of the animal's behavior, the results of the
present model are straightforward: the shapes of the optimal sensory
tuning curves should be adapted to that structure.

Two main conclusions follow from these results, a general one and a
specific one. The general observation is that, contrary to what is
implicitly assumed in most studies, the optimality of
sensory-triggered responses depends not only on their variability and
on the statistics of stimuli, but also on the downstream events driven
by those responses\rfs{MGLH05}.  If the downstream demands change, the
responses considered optimal will change as well, at least as required
by minimization of the performance measure used here, \EB. One
particular consequence of this is that the optimal width of peaked
tuning curves is not uniquely determined by signal-to-noise
considerations\rfd{ZS99}{BG06} (more on this below). This suggests
that a comprehensive understanding of the firing properties of sensory
neurons requires knowledge of the downstream impact of their
responses.

In retrospect, this point may seem obvious. If the motor functions to
be approximated are monotonic, so should be the tuning curves of
upstream neurons that drive them. However, this idea
has not been formally articulated before. Furthermore, previous
explanations of key features of tuning curves --- tuning curve width,
degree of overlap between curves, number of peaks, etc.\ --- have
always been based on arguments about coding efficiency. The simple
model presented here indicates that such features may also generally
depend on the motor actions performed by the animal. This, I believe,
is a new insight, because it applies to neurons that are firmly
considered as sensory.

The specific point is that monotonic and non-monotonic curves
are optimal under subtly different circumstances, which may depend on
what can be termed a `behavioral bias'. This simply refers to an
asymmetry in the relevant sensory stimulus. A bias exists when
different parts of the stimulus range lead to different sets of
possible actions, so that not all stimulus values are equal. The
classes of downstream functions used here were meant to abstract this
distinction in a simple way, and the results suggest that monotonic
curves are efficient when there is such an asymmetry. Image
contrast\rfs{AGFC02} and pressure on the skin\rfs{PSB00} are good
examples because, just on the basis of detection probability, high
values are much more likely to lead to behavioral responses than low
ones. But in general, weaker or more restricted biases may lead to
populations of neurons with both monotonic and peaked tuning curves,
as seen experimentally\rfr{HC05}{PvE05}. 

\subsec{Model Predictions}

If the model is correct, some variations in tuning properties across
sensory populations should correspond to adaptations that enhance
motor activity.  Specifically in the case of arrays of Gaussian tuning
curves\rfr{Para88}{BG06}, the model predicts that downstream motor
responses should vary more rapidly in the stimulus range where the
Gaussian curves are narrower.  For instance, according to Figure~4,
echolocating bats must compute motor functions that vary a lot around
zero echo delay. Elegant experimental studies by Moss and
collaborators are consistent with this interpretation.  They show not
only that the rate of turning of bats indeed increases as a target is
approached\rfd{GM06}{MBGS06}, as was argued earlier,
but also that their vocalizations speed up in several
ways: (1) the rate at which sonar calls are emitted increases as the
target gets near, (2) the duration of each call decreases, and (3)
each frequency-modulated call consists of a sweep from a high to a low
frequency, and the speed with which the frequencies are swept also
increases. These three quantities vary by a factor of about three from
the beginning to the end of a capture\rfs{MBGS06}. Furthermore, in the
brown bat, microstimulation of the superior colliculus produces not
only movements of the head and pinna but also sonar calls, where the
number of evoked sonar pulses increases as a function of both the
duration and the amplitude of the injected current\rfs{VSM02}. These
data strongly suggest that relevant motor neuron activity is indeed
generally faster in the region where narrow tuning curves are found.

The model may also be useful for understanding sensory responses
associated with escape or evasive behaviors in which, as a potential
threat approaches, the motor reaction should be faster. This is a
behavioral-bias scenario: if the likelihood or the speed of an evasive
movement increases monotonically as a function of the proximity and
speed of an incoming object, then one should expect the driving
sensory neurons to have monotonic profiles. This, indeed, is reported
to happen in several systems. For example, flying locusts make a
characteristic dive when predator-sized stimuli are looming on one
side. The key for triggering the glide is thought to be a single
movement detector unit, the so-called DCMD neuron, and this neuron
fires with increasing frequency as the looming stimulus gets
nearer\rfs{SRSS06}. Similar monotonic responses as functions of
distance\rfd{EHK99}{GN06} and speed\rfs{MJT91} have also been
documented in other neurophysiological preparations where escape or
collision avoidance is important. Even in monkeys, neurons that are
sensitive to the distance of an object approaching the face seem to
have monotonic dependencies on object distance (see Figure~4
in\rfs{GHG97}). Likewise, cortical neurons that respond to optic flow,
which are particularly useful for avoiding obstacles during
locomotion\rfs{GC06}, encode heading speed (i.e., the speed of one's
own motion) in a predominantly monotonic way\rfs{DW97}.

Perhaps the most counter-intuitive consequence of the model is that,
when behavior does not require high accuracy, the sensory
representation should be correspondingly coarse, even if, in
principle, it could be made more precise. As illustrated in Figure~2B
and 2F, when the motor response functions are broad, so should be the
sensory tuning curves. An impressive data set collected by Heffner and
colleagues supports this notion\rfd{Heff97}{Heff04}. They have shown
that sound localization capacity in mammals varies tremendously, with
discrimination thresholds ranging from about 1\deg\ in humans and
elephants to about 30\deg\ in mice and horses. These differences are
not accounted for by variations in interaural distance, animal
lifestyle, or environmental cues. Instead, ``sound localization acuity
in mammals appears to be a function of the precision required of the
visual orienting response to sound''\rfs{Heff97}. The argument is
this. A primary purpose of auditory localization is to generate an
orienting response, i.e., to bring the sound source into the fovea for
detailed visual analysis.  Consequently, species with small areas of
best vision (e.g., human, elephant) need to generate highly precise
movements, whereas species with large areas or streaks of best vision
(e.g., mouse, horse) do not.  Based on 24 mammalian species, the
correlation between sound localization acuity and foveal width is
0.92. Crucially, however, the correlation with visual acuity itself is
-0.31, so a purely sensory explanation again fails. According to
classic sensory coding notions, species with high acuity must have
tuning curves that are correspondingly narrower or less noisy.
Therefore, in view of the behavioral data, large variations in the
width of sound localization tuning curves are expected across species.
These would be explained by motor constraints, as predicted by the
theory.

Similar interpretations should be possible in other systems, as long
as sensory tuning curves can be directly related to clearly-defined
behaviors.

\subsec{Conclusion}

The model developed here is based on an optimality criterion for
neuronal tuning curves that takes into account both sensory (upstream)
and motor (downstream) processes. This simple model is useful when
sensory responses can be functionally related to specific behaviors,
in which case it may explain some features of sensory representations
that appear intriguing from the traditional perspective of sensory
coding based on information maximization. In particular, this approach
provides a theoretical rationale for the existence of monotonic tuning
curves, which so far have lacked a plausible explanation, and yields
some insight into the apparently idiosyncratic varieties of sensory
tuning curves observed across neurophysiological preparations.


\sec{Methods}
\vspace{-\baselineskip} 

\subsec{Numerical Methods}

All calculations were performed using Matlab (The Mathworks, Natick,
MA). Results are shown for $n$ between 2 and 8 neurons and $M\eq 50$
stimuli.  The mean response of neuron $i$ as a function of stimulus
$k$, where $k\eq 1,\ldots, 50$, was parameterized as follows
\b
   \avg{r_{ik}} = \frac{a_i \, \left( 1 + \exp\left(-h_i m_i\right) \right)^2}
             {\left( 1 + \exp\left(\left(-k + c_i - h_i\right) m_i\right) \, \right)
              \left( 1 + \exp\left(\left( k - c_i - h_i\right) m_i\right) \, \right)} 
   \label{tunc}
\e
where $a_i$ is the amplitude of the curve for neuron $i$, $c_i$ is the
center point, $h_i$ the half-width, and $m_i$ a factor that determines
the slope.  This expression produces either unimodal curves, which may
have positive or negative kurtosis, or monotonic curves, which may
vary in steepness. The correlation matrix $\mtx{C}$ was obtained by
assuming that noise is independent across neurons, in which case
$C_{ij} \eq \avg{r_{ik} \, r_{jk}} 
        \eq \avg{r_{ik}} \avg{r_{jk}} + \delta_{ij} \, \sigma^2_{jk}$,
where $\delta_{ij} \eq 1$ if $i\eq j$ and is zero otherwise. For each
neuron $i$, the SD of the noise during stimulus $k$ was
$\sigma_{ik} \eq \alpha \left( r_{\mathrm{max}}/2 + 
                               \sqrt{\avg{r_{ik}}} \right)$,
but other choices produced similar results. In the low- and high-noise
conditions, $\alpha \eq 0.05$ and 0.5, respectively.

Matrices \FI\ were produced directly by generating 5000 functions
$\vec{F}_{\alpha}$ within a class and averaging the pairwise products
$F_{\alpha k} F_{\alpha l}$.  The functions
in each class were determined by small numbers of parameters. For
example, for the saturating monotonic curves (Figure~2C), 
$F_{\alpha k} = a + 
     b \left(1 + \exp\left(\left(c - k\right)/d\right)\right)^{-1}$,
where, for each $\alpha$, the center point $c$ and slope factor $d$
were chosen randomly within a range and $a$ and $b$ were set to
satisfy two normalization conditions. The first one was 
$\sum_{k=1}^M s_k \, F_{\alpha k} \eq 0$ 
for all $\alpha$, so the mean of each downstream function was set
to zero. This was simply to shift the baseline of each
$\vec{F}_{\alpha}$ and make the resulting \FI\ matrix easier to
visualize in the plots; it had little or no effect on the optimal
tuning curves. The second normalization condition was
$\sum_{k=1}^M s_k \, \Phi_{kk} \eq 1$.
It limited the amplitude of the downstream responses. Final values
of $\Phi_{kl}$ varied depending on the chosen class of functions, 
but never exceeded the range $[-2.4, 4]$.

Given the terms $s_k$ and $\Phi_{kl}$, a routine searched for the
combinations of parameters $a$, $h$, $c$ and $m$ in
Equation~\ref{tunc} that minimized \EB\ (Equation~\ref{EBerr}). The
minimization routine used the Nelder-Mead downhill simplex
method\rfs{PFTV92}. A set of tuning curves was deemed optimal only
after extensive testing and refining to ensure that the solution was
near the global minimum.  Additional constraints were included by
adding suitable penalty terms to \EB.  For instance, to constrain the
total power, a term proportional to 
$\sum_{i=1}^n \sum_{k=1}^M s_k\, r_{ik}^2$
was added.

Tuning curves were also generated using a second parameterization
(Figure~3 and Supporting Information). In this case,
Equation~\ref{tunc} was substituted with a combination of two
half-Gaussians with different widths and baselines but a common
peak\rfs{HC05},
\b
   \avg{r_{ik}} = \left\{ \begin{array}{cc}     
                  b_i^{-} + \left(a_i - b_i^{-}\right)
                       \exp\!\left(-(k-c_i)^2/(h^-_i)^2 \right)
                  & \mathrm{if\ } k \leq c_i \rule[-13pt]{0pt}{1pt} \\
                  b_i^{+} + \left(a_i - b_i^{+}\right)
                       \exp\!\left(-(k-c_i)^2/(h^+_i)^2 \right)
                  & \mathrm{if\ } k > c_i 
                 \end{array} \right. \, .
   \label{tunc1}
\e
Here there are six free parameters per basis neuron: the center point
$c_i$, amplitude $a_i$, left and right baseline levels $b_i^{-}$ and
$b_i^{+}$, and left and right widths $h_i^{-}$ and $h_i^{+}$.

\subsec{Derivation of \EB}

The objective here is to derive an expression that quantifies how well
the sensory tuning curves $\avg{\mtx{r}}$ approximate a desired set of
downstream responses \FF. The standard procedure is to consider the
average squared difference between driven and desired responses, i.e.,
the norm 
$\left|\mtx{R} - \mtx{F}\right|$. Because
$\mtx{R} \eq \mtx{w} \, \mtx{r}$, this produces
\b
     E_L = \frac{1}{N} \, \sum_{\alpha=1}^{N} \sum_{k=1}^{M}
           s_k \left< \left(
                   \sum_{i=1}^n w_{\alpha i} \, r_{ik} - F_{\alpha k} 
               \right)^{\!\!2} \, \right> 
     \label{ELerr}
\e 
where the coefficient $s_k$ represents the probability of stimulus
$k$, such that $\sum_{k=1}^M s_k \eq 1$, and the average indicated by
angle brackets is over repeated presentations of a given stimulus.
\EL\ is the linear approximation error. This number quantifies how
accurately the sensory (basis) neurons and associated synaptic weights
are able to generate the desired motor activity downstream.  

The next step is to obtain an expression for the error that no longer
depends on the synaptic connections. To do this, the idea is to find
the set of synaptic weights $\mtx{w}^{\mathrm{opt}}$ that minimize
\EL\ assuming that the sensory tuning curves are known. These optimal
weights are then substituted back into Equation~\ref{ELerr}, and the
result is an expression for the mean square error that assumes that
the synaptic connections are always the best possible ones. This is as
follows.

First, find the optimal connections by calculating the partial
derivatives of \EL\ in Equation~\ref{ELerr} with respect to 
$w_{\alpha i}$ and equating the result to zero. This gives the optimal
weights 
\b
   w_{\alpha i}^{\mathrm{opt}} = \sum_{k=1}^M \sum_{j=1}^n 
                         s_k \, F_{\alpha k} \avg{r_{jk}} C^{-1}_{ji}
   \label{wopt}
\e
where $\mtx{C^{-1}}$ is the inverse of $\mtx{C}$, and
$C_{ij} \eq \sum_{k=1}^M s_k \, \avg{r_{ik} \, r_{jk}}$ 
is the correlation between sensory neurons $i$ and $j$. The weights
$\mtx{w}^{\mathrm{opt}}$ generate linearly-driven responses $\mtx{R}$
that, on average, approximate the target motor responses as accurately
as possible given the mean firing rates of the sensory neurons and the
statistics of the stimuli. 

Having minimized \EL\ with respect to the connections, next, find out
how big it is by substituting the optimal weights of
Equation~\ref{wopt} back into Equation~\ref{ELerr} and rearranging
terms. Calling the result \EB, this gives 
\b
    E_B = 1 - \sum_{i=1}^n \sum_{j=1}^n Q_{ij} \, C^{-1}_{ij} 
    \label{EBerr}
\e
where
\b
    Q_{ij} = \sum_{k=1}^M \sum_{l=1}^M s_k \, s_l 
                         \avg{r_{ik}} \Phi_{kl} \avg{r_{jl}} 
    \label{Qmtx} 
\e
and 
$\Phi_{kl} = \frac{1}{N} \, 
             \sum_{\alpha=1}^N F_{\alpha k} \, F_{\alpha l}$, 
as mentioned in the main text. Thus, \EB\ is a function of the 
first and second moments of the sensory responses, the stimulus
probabilities, and the output correlations \FI. 

Importantly, note that in the expression above, the following
normalization condition was imposed:
$\sum_{k=1}^M s_k \, \Phi_{kk} \eq 1$.
This limits the amplitude of the downstream functions and bounds \EB\
between 0 and 1. That the error cannot be negative follows directly
from the definition of \EL\ above. That it is bounded by 1 is not
immediately obvious, but is a consequence of the fact that the
eigenvalues of \FI\ are non-negative, and the normalization restricts
their total sum. See the next section for more details.

\subsec{Lower Bounds on the Approximation Error}

In this section, the two analytic results discussed in the main text,
Equations~\ref{EBerr1} and \ref{EBerr2}, are developed. The main
expression derived below is, in fact, a slightly more general
statement about the accuracy of the sensory tuning curves. 

Here, a key simplifying assumption is that all stimulus probabilities
are equal, so $s_k \eq 1/M$ for all $k$. As a consequence, the maximum
eigenvalue of \FI\ satisfies 
$1 \leq \lambda^{\mathrm{max}} \leq M$.
This important property is true for two reasons, first, because \FI\
results from the product of a matrix times its transpose, which
guarantees that all its eigenvalues are non-negative, and second,
because of the normalization condition on \FI, which in this case is
\b
   \sum_{k=1}^M s_k \, \Phi_{kk} = 
                    \frac{1}{M} \, \sum_{k=1}^M \lambda_{k} = 1 .
   \label{FInorm}
\e
This condition makes the sum over all eigenvalues equal to $M$. Hence,
$\lambda^{\mathrm{max}}$ is bounded between 1 and $M$.

To see how \EB\ depends on the sensory responses, recall that their
correlation matrix $\mtx{C}$ is such that
$C_{ij} \eq \sum_{k=1}^M s_k \, \avg{r_{ik} \, r_{jk}}$. 
Then, for a single basis neuron ($n \eq 1$) with mean response
$\avg{r_k}$, $\mtx{C}$ becomes a scalar
$C \eq \ol{r}^2 + \ol{\sigma}^2$, where
\b
    \ol{r}^2      = \frac{1}{M} \sum_{k=1}^{M} 
                            \avg{r_k}^2 , \hspace{1cm}
    \ol{\sigma}^2 = \frac{1}{M} \sum_{k=1}^{M} 
                            \avg{ \left(r_k - \avg{r_k}\right)^2 } .
  \label{avgr2} \\
\e
Also, if the tuning curve is an eigenvector of \FI\ with eigenvalue
$\lambda$, then
$\sum_{l=1}^M \Phi_{kl} \avg{r_l} \eq \lambda \avg{r_k}$,
by definition, and Equation~\ref{Qmtx} gives
$Q \eq \lambda \, \ol{r}^2 / M$. 
Substituting into Equation~\ref{EBerr} and defining 
$\rho \eq \ol{r}^2 / \ol{\sigma}^2$
leads to Equation~\ref{EBerr1}, which is the approximation error for a
single neuron.

With more neurons, it is possible to derive a lower bound on the error
that is more general than Equation~\ref{EBerr2}.  First, assume that 
$s_k\eq 1/M$ and that the noise has equal magnitude and is
uncorrelated across neurons, such that
\ba
    \mtx{C} & = & \frac{1}{M} \avg{\mtx{r} \mtx{r}\tsp}  
            \; = \; \frac{1}{M} \avg{\mtx{r}} \avg{\mtx{r}}\tsp
                    + \ol{\sigma}^2 \, \mtx{I}_{n} 
      \nonumber \\
    \mtx{Q} & = & \frac{1}{M^2} \, 
                  \avg{\mtx{r}} \mtx{\Phi} \avg{\mtx{r}}\tsp
      \label{QandC}
\ea
where $\mtx{I}_{n}$ is the \nn\ identity matrix and $\ol{\sigma}^2$,
the variance averaged over stimuli, is the same for all neurons. To
proceed, consider the singular value decomposition (SVD) of the matrix
of mean responses, 
$\avg{\mtx{r}} = \mtx{u} \mtx{S} \mtx{V}\tsp$, 
where $\mtx{u}$ is an \nn\ unitary matrix, $\mtx{V}$ is an \MM\
unitary matrix, and $\mtx{S}$ is an \nM\ matrix with $n$ singular
values along the diagonal and zeros elsewhere\rfs{PFTV92}. This is
assuming that the $n$ tuning curves are independent; if not, then the
number of non-zero elements of $\mtx{S}$ will equal the number of
independent curves (the rank of $\avg{\mtx{r}}$). The SVD is a
generalization to rectangular matrices of the classic eigenvalue
decomposition.  Substituting $\mtx{C}$ and $\mtx{Q}$ into
Equation~\ref{EBerr} and using the SVD representation of
$\avg{\mtx{r}}$ leads to
\b
    E_B = 1 - \frac{1}{M} \, \sum_{j=1}^{M} \left( 
             \mtx{V}\tsp \mtx{\Phi} \mtx{V} \,
             \mtx{S}\tsp \left( \mtx{S} \mtx{S}\tsp  
                  + M \ol{\sigma}^2 \, \mtx{I}_{n} \right)^{-1} 
             \mtx{S} \right)_{\!\!jj}
        = 1 - \frac{1}{M} \, \sum_{j=1}^{M} \left( 
          \mtx{V}\tsp \mtx{\Phi} \mtx{V} \, \mtx{D} \right)_{jj} .
    \label{errSVD}
\e
The first equality results from the defining property of unitary
matrices, such that
$\mtx{u} \mtx{u}\tsp \eq \mtx{I}_n$ 
and 
$\mtx{V} \mtx{V}\tsp \eq \mtx{I}_M$. 
The second equality results from grouping into $\mtx{D}$ all the terms
involving $\mtx{S}$. The matrix $\mtx{D}$ turns out to be \MM\ and
diagonal, with entries 
$D_{i} \eq S_{i}^2/(S_{i}^2 + M \ol{\sigma}^2)$,
where a single index is used to indicate relevant elements in diagonal
matrices. Note, however, that only the first $n$ diagonal elements are
non-zero, because $\mtx{S}$ itself only has at most $n$ non-zero
singular values along the diagonal (recall that $\mtx{S}$ is diagonal
but rectangular, \nM).  The lower bound on the expression above thus
involves a sum of only $n$ terms; the bound is
\b
    E_B \geq 1 - \frac{1}{M} \: \sum_{i=1}^{n} 
             \lambda_i \; \frac{S_{i}^2}{S_{i}^2 + M \ol{\sigma}^2}
    \label{errSVD1}
\e
where $\lambda_1, \ldots, \lambda_n$ are the $n$ largest eigenvalues
of \FI. 

To see this, first write \FI\ in terms of its eigenvalue
decomposition, so that
$\mtx{V}\tsp \mtx{\Phi} \mtx{V} \eq 
\mtx{V}\tsp \mtx{E} \mtx{\Lambda} \mtx{E}\tsp \mtx{V}$,
where $\mtx{E}$ is the matrix of (right) eigenvectors of \FI. Suppose
that the eigenvalues $\mtx{\Lambda}$ are sorted in decreasing order,
so that $\lambda_1$ is the largest. Now note that the diagonal
elements of the matrix 
$\mtx{V}\tsp \mtx{\Phi} \mtx{V}$
depend on the match between $\mtx{V}$ and $\mtx{E}$. In particular,
the best possible match occurs when $\mtx{V}$ is identical to
$\mtx{E}$; then   
$\mtx{V}\tsp \mtx{E} \, \mtx{\Lambda} \,  
                        \mtx{E}\tsp \mtx{V} \eq  \mtx{\Lambda}$
and the equality in Equation~\ref{errSVD1} follows directly from
Equation~\ref{errSVD}. This means that equality is obtained when the
basis tuning curves are constructed using the eigenvectors of \FI\
sorted in decreasing order (i.e., $\mtx{V}$ is equal to $\mtx{E}$).
In contrast, if, for example, $\mtx{V}$ has the same columns as
$\mtx{E}$ but sorted in the reverse order, then the resulting sum is
similar to that in Equation~\ref{errSVD1} except that it involves the
$n$ smallest eigenvalues. That \EB\ varies between 0 and 1 follows
from Equation~\ref{FInorm}.

Equation~\ref{errSVD1} is the main analytic result and provides
important intuitions about the mean basis responses, or sensory tuning
curves, $\avg{\mtx{r}}$. These are as follows.
\begin{enumerate}
\item
With $n \eq 1$ the result is Equation~\ref{EBerr1} written as an
inequality, with the signal-to-noise ratio equal to 
$S^2/(M\ol{\sigma}^2)$.
Also, Equation~\ref{EBerr2} is obtained when $\ol{\sigma}^2 \eq 0$. 
\item
Noise always increases the error, because $\ol{\sigma}^2$ effectively
decreases every eigenvalue in Equation~\ref{errSVD1}.
\item
Noise partially determines the optimal shapes of the tuning curves.
For example, if $\lambda_1 > \lambda_2$ but $S_2 \gg S_1$, then the
second eigenvector should take precedence over the first, because its
signal-to-noise ratio will be much higher. In other words, in this
case the first column in $\mtx{V}$ should contain the second
eigenvector of \FI. Thus, noise also determines which eigenvectors
should be chosen in what order, and therefore the optimal shapes of
the basis responses. 
\item
Because of the last point, noise helps solve the ambiguity discussed
earlier --- that the set of basis responses is determined up to an
invertible transformation.  However, it does not entirely solve the
problem, and this is why.  When $\ol{\sigma} \eq 0$, both matrices
$\mtx{u}$ and $\mtx{S}$ are absent from Equation~\ref{errSVD1}.
Therefore, they are arbitrary; they do not affect the error (as long
as they are unitary and diagonal, as required).  In contrast, with
noise there is a criterion for setting the $S_i$ values, so only
$\mtx{u}$ remains arbitrary. Thus, without noise $\avg{\mtx{r}}$ is
ambiguous up to an invertible transformation, whereas with noise it is
ambiguous up to a unitary transformation. 
\item
In addition, it is important to mention that this ambiguity remains
when the stimulus probabilities are not uniform. When arbitrary
probability values $s_k$ are included in the calculation of
Equation~\ref{errSVD}, the resulting expression for the error still
does not depend on $\mtx{u}$.  Therefore, manipulating the stimulus
statistics does not solve this problem.  
\item
Finally, to minimize \EB\ in the presence of noise it is best to
increase $S_i^2$ as much as possible. However, the total power in the
mean responses is 
$\sum_{i=1}^n \sum_{k=1}^M \avg{r_{ik}}^2 = \sum_{i=1}^n S_i^2$.  
Therefore, with noise, additional constraints that effectively limit
the power are necessary to obtain optimal responses of finite
amplitude.
\end{enumerate}


\sec{Supporting Information}

Supporting information is appended to this preprint. It is also found
online at DOI: 10.1371/journal.pbio.0040387.sd001 (154 KB PDF).

\sec{Acknowledgments}

I thank Terry Stanford for useful discussions and Peter Latham for
many valuable suggestions. Research was partially supported by the
National Institute for Neurological Disorders and Stroke grant
NS044894.  


\sec{References}

\begin{refs}

\item \label{Atic92}
Atick JJ (1992) 
Could information theory provide an ecological theory of sensory processing? 
Network 3: 213--251.  

\item \label{Barl01}
Barlow H (2001)
Redundancy reduction revisited.
Network 12: 241--253.

\item \label{Simo03}
Simoncelli EP (2003)
Vision and the statistics of the visual environment.
Curr Opin Neurobiol 13: 144--149. 

\item \label{AR90}
Atick JJ, Redlich AN (1990) 
Towards a theory of early visual processing.
Neural Comput.\ 2: 308--320.  

\item \label{AR92}
Atick JJ, Redlich AN (1992) 
What does the retina know about natural scenes? 
Neural Comput 4: 196--210.  
 
\item \label{OF96}
Olshausen BA, Field DJ (1996) 
Emergence of simple-cell receptive field properties by learning a sparse 
code for natural images. 
Nature 381: 607--609.

\item \label{BS97}
Bell AJ, Sejnowski TJ (1997)
The `independent components' of natural scenes are edge filters.
Vision Res 37: 3327--3338. 

\item \label{Lewi02}
Lewicki MS (2002)
Efficient coding of natural sounds.
Nat Neurosci 5: 356--363. 

\item \label{SL06}
Smith EC, Lewicki MS (2006)
Efficient auditory coding.
Nature 439: 978--982. 
 
\item \label{DAR96}
Dan Y, Atick JJ, Reid C (1996) 
Efficient coding of natural scenes in the lateral geniculate nucleus: 
experimental test of a computational theory.
J Neurosci 16(10): 3351--3362.

\item \label{VG00}
Vinje WE, Gallant JL (2000)
Sparse coding and decorrelation in primary visual cortex 
during natural vision.
Science 287: 1273--1276.

\item \label{SS01}
Schwartz O, Simoncelli EP (2001) 
Natural signal statistics and sensory gain control. 
Nat Neurosci 4: 819--825.

\item \label{CWT04}
Caywood MS, Willmore B, Tolhurst DJ (2004)
Independent components of color natural scenes resemble V1 neurons in 
their spatial and color tuning.
J Neurophysiol 91: 2859--2873. 

\item \label{MGLH05}
Machens CK, Gollisch T, Kolesnikova O, Herz AV (2005) 
Testing the efficiency of sensory coding with optimal stimulus ensembles. 
Neuron 47: 447--456. 

\item \label{MSFKSRH03}
Machens CK, Sch\"utze H, Franz A, Kolesnikova O, Stemmler MB,
Ronacher B, Herz AV (2003) 
Single auditory neurons rapidly discriminate conspecific communication
signals.
Nat Neurosci 6: 341--342.

\item \label{Albr84}
Albright TD (1984)
Direction and orientation selectivity of neurons in visual area 
MT of the macaque.
J Neurophysiol 52: 1106--1130. 

\item \label{FBG-R89}
Funahashi S, Bruce CJ, Goldman-Rakic PS (1989) 
Mnemonic coding of visual space in the monkeys's dorsolateral 
prefrontal cortex. 
J Neurophysiol 61(2): 331--349.  

\item \label{TMR90}
Taube JS, Muller RU, Ranck JB Jr (1990)
Head-direction cells recorded from the postsubiculum in freely moving rats. 
I. Description and quantitative analysis.
J Neurosci 10: 420--435. 
 
\item \label{MJT91}
Miller JP, Jacobs GA, Theunissen F (1991) 
Representation of sensory information in the cricket cercal sensory 
system. I. Response properties of the primary interneurons. 
J Neurophysiol 66: 1680--1689.  

\item \label{O'KB96}
O'Keefe J, Burgess N (1996)
Geometric determinants of the place fields of hippocampal neurons.
Nature 381: 425--428. 
 

\item \label{Para88}
Paradiso MA (1988) 
A theory for the use of visual orientation information which exploits 
the columnar structure of striate cortex.
Biol Cybern 58: 35--49.  

\item \label{SA94}
Salinas E, Abbott LF (1994) 
Vector reconstruction from firing rates. 
J Comput Neurosci 1: 89--107.  

\item \label{ZS99}
Zhang K, Sejnowski TJ (1999)
Neuronal tuning: To sharpen or broaden?
Neural Comput 11: 75--84. 

\item \label{DLP99}
Deneve S, Latham PE, Pouget A (1999)
Reading population codes: a neural implementation of ideal observers.
Nat Neurosci 2: 740--745.

\item \label{BG06}
Butts DA, Goldman MS (2006)
Tuning curves, neuronal variability and sensory coding.
PLoS Biol 4(4): e92.

\item \label{RBHL99}
Romo R, Brody CD, Hern\'andez A, Lemus L (1999)
Neuronal correlates of parametric working memory in the prefrontal cortex. 
Nature 399: 470--473.

\item \label{PSB00}
Pruett JR, Sinclair RJ, Burton H (2000)
Response patterns in second somatosensory cortex (SII) of
awake monkeys to passively applied tactile gratings.
J Neurophysiol 84: 780--797.

\item \label{SHZR00}
Salinas E, Hern\'andez H, Zainos A, Romo R  (2000)
Periodicity and firing rate as candidate neural codes for 
the frequency of vibrotactile stimuli.
J Neurosci 20: 5503--5515.

\item \label{BITDH97}
Bremmer F, Ilg UJ, Thiele A, Distler C, Hoffmann KP (1997)
Eye position effects in monkey cortex. I. Visual and pursuit-related activity 
in extrastriate areas MT and MST.
J Neurophysiol 77: 944--961. 

\item \label{AGFC02}
Albrecht DG, Geisler WS, Frazor RA, Crane AM (2002)
Visual cortex neurons of monkeys and cats: temporal dynamics of the 
contrast response function.
J Neurophysiol 88: 888--913. 

\item \label{KBOdBV05}
Kayaert G, Biederman I, Op de Beeck HP, Vogels R (2005)
Tuning for shape dimensions in macaque inferior temporal cortex.
Eur J Neurosci 22: 212--224. 

\item \label{Guig03}
Guigon E (2003)
Computing with populations of monotonically tuned neurons.
Neural Comput 15: 2115--2127. 

\item \label{HC05}
Hinkle DA, Connor CE (2005)  
Quantitative characterization of disparity tuning in ventral pathway area V4.
J Neurophysiol 94: 2726--2737.  

\item \label{ZHB04}
Zhang T, Heuer HW, Britten KH (2004)
Parietal area VIP neuronal responses to heading stimuli are encoded 
in head-centered coordinates.
Neuron 42: 993--1001 .

\item \label{PvE05}
Peng X, Van Essen DC (2005)
Peaked encoding of relative luminance in macaque areas V1 and V2.
J Neurophysiol 93: 1620--1632. 

\item \label{Pogg90}
Poggio T (1990)
A theory of how the brain might work. 
Cold Spring Harbor Symp Quant Biol 5: 899--910.

\item \label{PS97}
Pouget A, Sejnowski TJ (1997)
Spatial tranformations in the parietal cortex using basis functions.
J Cog Neurosci 9: 222--237.

\item \label{PG89}
Poggio T, Girosi F (1989)
A Theory of Networks for Approximation and Learning.
AI Memo 1140, Massachusetts Institute of Technology. 

\item \label{SA00}
Salinas E and Abbott LF (2000) 
Do simple cells in primary visual cortex form a tight frame? 
Neural Computation 12: 313-335.

\item \label{LS99}
Lee DD, Seung HS (1999)  
Learning the parts of objects by non-negative matrix factorization.
Nature 401: 788--791. 

\item \label{Joll02}
Jollife (2002) 
Principal Component Analysis.
New York: Springer-Verlag.

\item \label{GM06}
Ghose K, Moss CF (2006)
Steering by hearing: a bat's acoustic gaze is linked to its flight 
motor output by a delayed, adaptive linear law.
J Neurosci 26: 1704--1710. 

\item \label{MBGS06}
Moss CF, Bohn K, Gilkenson H, Surlykke A (2006) 
Active listening for spatial orientation in a complex auditory scene. 
PLoS Biol 4(4): e79.

\item \label{SH86}
Suga N, Horikawa J (1986) 
Multiple time axes for representation of echo delays in the auditory 
cortex of the mustached bat. 
J Neurophysiol 55: 776--805.  

\item \label{OS91}
Olsen JF, Suga N (1991)
Combination-sensitive neurons in the medial geniculate body of the 
mustached bat: encoding of relative velocity information. 
J Neurophysiol 65: 1254--1274.

\item \label{Adol93}
Adolphs R (1993)
Bilateral inhibition generates neuronal responses tuned to interaural 
level differences in the auditory brainstem of the barn owl.
J Neurosci 9: 3647--3668. 

\item \label{VSM02}
Valentine DE, Sinha SR, Moss CF (2002)
Orienting responses and vocalizations produced by microstimulation in the 
superior colliculus of the echolocating bat, Eptesicus fuscus.
J Comp Physiol A Neuroethol Sens Neural Behav Physiol 188: 89--108.  

\item \label{SRSS06}
Santer RD, Rind FC, Stafford R, Simmons PJ (2006)
Role of an identified looming-sensitive neuron in triggering 
a flying locust's escape.
J Neurophysiol 95: 3391--3400. 

\item \label{EHK99}
Edwards DH, Heitler WJ, Krasne FB (1999)
Fifty years of a command neuron: the neurobiology of escape behavior 
in the crayfish. 
Trends Neurosci 22: 153–-161.

\item \label{GN06}
Gallagher SP, Northmore DP (2006)
Responses of the teleostean nucleus isthmi to looming objects 
and other moving stimuli.
Vis Neurosci 23: 209--219. 

\item \label{GHG97}
Graziano MSA, Hu TX, Gross CG (1997)
Visuospatial properties of ventral premotor cortex.
J Neurophysiol 77: 2268--2292.

\item \label{GC06}
Graziano MS, Cooke DF (2006)
Parieto-frontal interactions, personal space, and defensive behavior.
Neuropsychologia 44: 845--859.  

\item \label{DW97}
Duffy CJ, Wurtz RH (1997)  
Medial superior temporal area neurons respond to speed patterns 
in optic flow.
J Neurosci 17: 2839--2851. 
 
\item \label{Heff97}
Heffner RS (1997)
Comparative study of sound localization and its anatomical 
correlates in mammals. 
Acta Otolaryngological (Stockholm), Suppl 532: 46--53.

\item \label{Heff04}
Heffner RS (2004) 
Primate hearing from a mammalian perspective. 
The Anatomical Record Part A, 281A: 1111--1122.

\item \label{PFTV92}
Press WH, Flannery BP, Teukolsky SA, Vetterling WT (1992)
Numerical Recipes in C. 
New York: Cambridge University Press.  

\end{refs}

\clearpage


\pagenumbering{arabic}

\begin{Large}
\noindent
{\sf \textbf{Supporting Information \\ }}
\end{Large}

\noindent
Main text:
Salinas E (2006) How behavioral constraints may determine
optimal sensory representations. PLoS Biology 4(12): e387\@.
DOI: 10.1371/journal.pbio.0040387.
\vspace*{0.1in} 

\sec{Parameter Manipulations}

Figure S1 shows additional results in which the same four classes of
target functions shown in Figure~2 were used but other aspects of the
computer experiments were varied.  In Figure~S1A, the noise of the
basis responses was much higher.  
\begin{figure*}[b!]
\centerline{\epsfig{figure=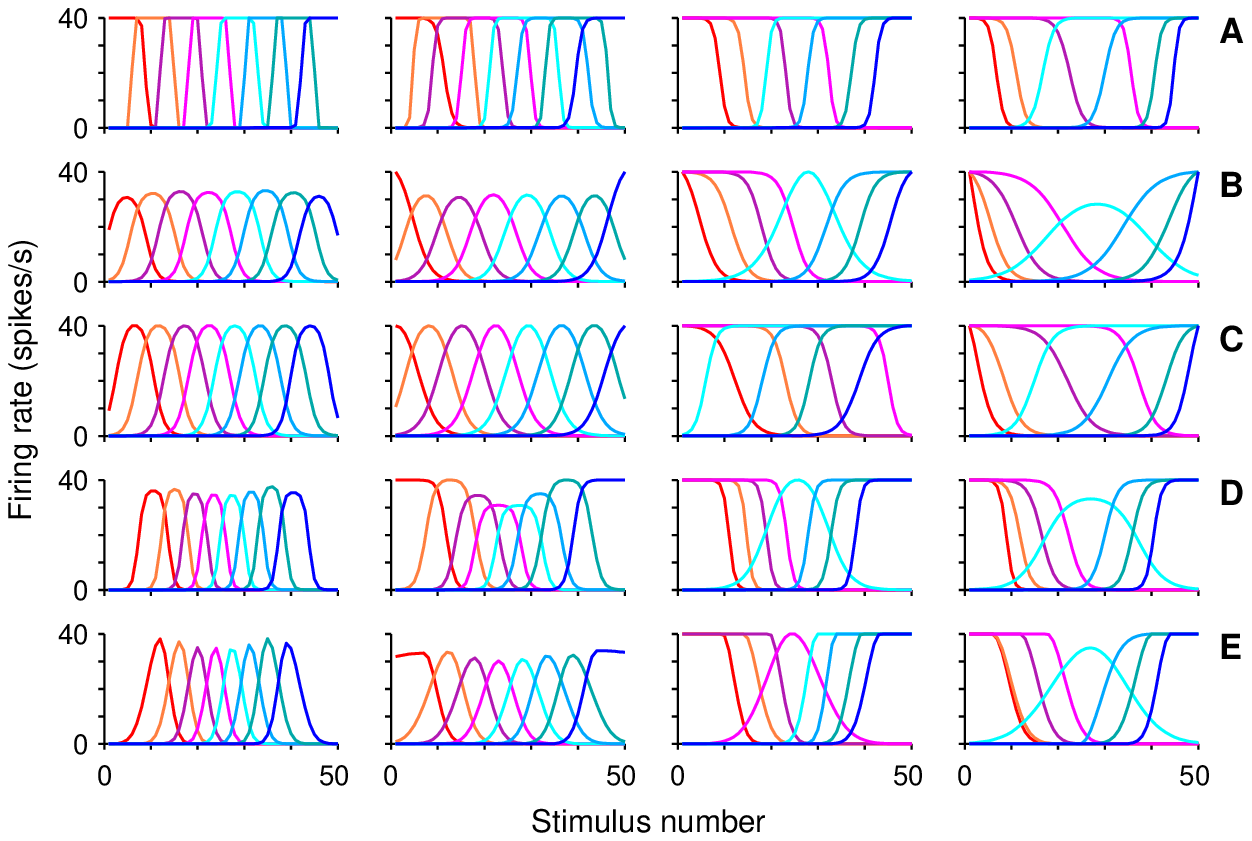,width=4.6in,clip} }
\vspace{6pt}
\parbox[t]{\textwidth}{\small
{\sf \textbf{Figure S1\@.}} 
Sets of eight optimal tuning curves obtained under various
conditions.
Columns correspond to the same classes of downstream functions as in
Figure~2A--D.  Numerical experiments were as in Figure~2, except for
changes stated explicitly.
\textsf{(A)} High noise. The SD of the firing rates was ten times
as high as in Figure~2. 
\textsf{(B)} High power cost. A penalty term proportional to the sum
of the squared responses was added to \EB. 
\textsf{(C)} Unequal stimulus probabilities.  The probability profile
($s_k$ as a function of $k$) was Gaussian with an SD of 7. The
stimulus in the middle of the range was about ten times more frequent
than those at the edges. 
\textsf{(D)} Combined conditions. The three previous manipulations
were applied simultaneously.
\textsf{(E)} As in \textsf{(D)}, but with the tuning curves
parameterized in an alternative way (Equation~\ref{tunc1}).  }
\end{figure*}
This caused all tuning curves to rise and fall more steeply than with
low noise. This makes sense because, to increase the signal-to-noise
ratio as much as possible, higher firing rates are necessary. In
Figure~S1B, a term penalizing the total power of the responses was
added to the error (see Methods), which amounts to putting an energy
cost on the neural activity. This tends to favor unimodal neurons,
which typically have smaller mean firing rates across all stimuli than
monotonic ones. In this case, unimodal curves reduced their amplitudes
and monotonic curves shifted toward the edges. Also, a monotonic curve
was exchanged for a unimodal one.  Such exchanges occur only when
adding a neuron results in a large power penalty, larger than the
corresponding decrease in \EB.  In the examples shown, \EB\ was
already very low with 7 neurons, so the advantage in accuracy of an
additional monotonic curve was too small relative to its high mean
rate.  In Figure~S1C, the stimulus probabilities were not uniform.
Instead, they had a Gaussian profile centered on the middle stimulus
(see Figure~S4F).  This caused the center points of the tuning curves
to shift very slightly toward the high-frequency stimuli. The effect,
however, was much stronger in conjunction with the other
manipulations, as shown in Figure~S1D\@. Here, all three changes ---
high noise, high power cost, and unequal stimulus probabilities ---
were applied simultaneously.  The tuning curve locations varied
strongly but their shapes remained qualitatively the same. Finally, to
investigate whether the results depended on the specific tuning curve
parameterization that was chosen, all numerical experiments were
repeated with a second parameterization that allowed tuning curve
profiles intermediate between unimodal and monotonic; this was the
same parameterization used by Hinkle and Connor\rfs{HC05} (see Methods
and Figure~3B).  In all cases, the results were similar to those
obtained earlier. Figure~S1E shows an example in which the same
conditions of Figure~S1D were used.  Although the individual shapes
are slightly different, as expected, all curves are either unimodal or
monotonic --- no intermediate curves were generated.  

In summary, then, the results are not overly sensitive to noise, power
constraints, stimulus probabilities, or the specific way in which
tuning curves are defined mathematically.

\sec{An Alternative Set of Constraints}

A different approach was also explored in which Equation~\ref{ELerr}
was directly minimized with respect to $\mtx{w}$ and $\avg{\mtx{r}}$
using a modified gradient descent algorithm. In this case, both
$\mtx{w}$ and $\avg{\mtx{r}}$ were updated iteratively. Tuning curves
were constrained to vary between 0 and 40, which was enforced by
renormalizing all modified curves after every update. In addition, the
synaptic connections were forced to be sparse by adding to
Equation~\ref{ELerr} a penalty term proportional to
\b
   \sum_{\alpha}^N \sum_{i,j\neq i}^n 
                      \left| w_{\alpha i} \, w_{\alpha j} \right|
   + \sum_{i}^n \sum_{\alpha,\beta \neq \alpha}^N 
                      \left| w_{\alpha i} \, w_{\beta i} \right| .
   \label{redun}
\e
Here, the first term represents the synaptic redundancy across rows of
$\mtx{w}$, whereas the second term represents the redundancy across
columns. For instance, the penalty term for row $\alpha$ is 
$\sum_{i,j\neq i}^n \left| w_{\alpha i} \, w_{\alpha j} \right|$. 
This expression is minimized when only one of the connections to
downstream neuron $\alpha$ is not zero, in which case the actual value
of the non-zero weight does not matter. This type of penalty thus tends
to produce many synaptic weights equal to zero. When the number of
basis neurons is equal to the number of downstream neurons, this
constraint assigns one basis neuron to one downstream neuron (if no
other restrictions are imposed), making the connectivity matrix
equivalent to the unit matrix. Thus, an intuitive interpretation of
this constraint is ``construct each downstream response using as few
basis neurons as possible''.

The results that follow were obtained by minimizing \EL\
(Equation~\ref{ELerr}) plus the expression above multiplied by a
constant that determined the penalty strength.  To use as few
restrictions as possible, the motor responses $\mtx{F}$ and the
sensory tuning curves were made to vary between 0 and 40; no other
normalization conditions were invoked.

\begin{figure*}[t!]
\centerline{\epsfig{figure=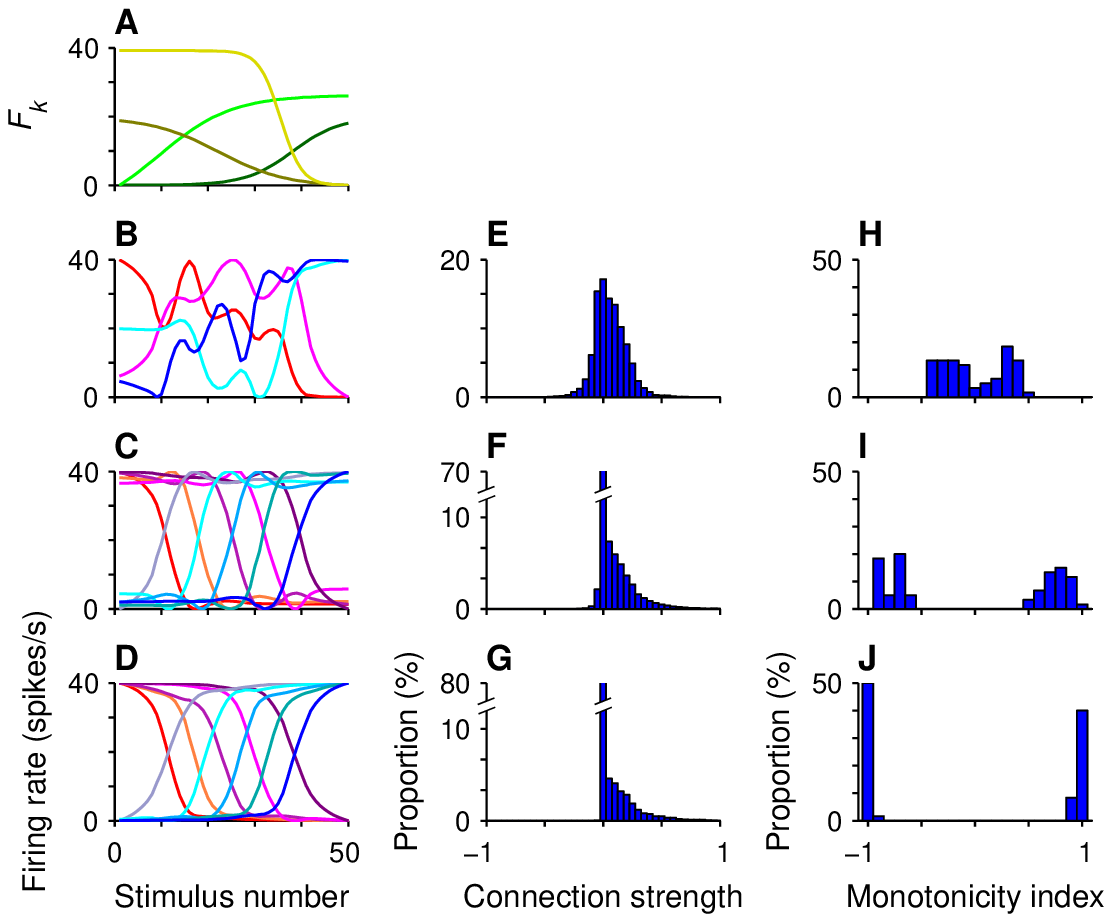,width=3.5in,clip} \hspace{0.10in}
\parbox[b]{2.40in}{\small
{\sf \textbf{Figure S2.}} 
Optimal tuning with sparse connectivity.
Results are shown for \FF\ with 2000 downstream functions and sets of
10 basis functions.
\textsf{(A)} Four examples of desired downstream responses.
\textsf{(B)} Four examples of optimal tuning curves obtained with no
connectivity constraint (penalty strength was zero). 
\textsf{(C, D)} Optimal tuning curves obtained with penalty strengths
of 0.002 and 0.02, respectively.
\textsf{(E--G)} Distributions of synaptic weight values for penalty
strengths of 0, 0.002 and 0.02, respectively.
\textsf{(H--J)} Distributions of monotonicity indices for penalty
strengths of 0, 0.002 and 0.02, respectively. \vspace{2pt}}}
\end{figure*}

Figure S2 illustrates the effect of the sparse connectivity
constraint. Here, \FF\ consisted of increasing and decreasing
sigmoidal functions, as in Figure~2C\@. Without the sparseness
constraint, the distribution of optimal connections that results is
approximately normal (Figure~S2E), and the corresponding optimal basis
functions have multiple peaks and no particular structure
(Figure~S2B). This is because, as mentioned in the main text, without
additional constraints on $\mtx{w}$ or $\avg{\mtx{r}}$ the
minimization of \EL\ is an ill-posed problem; there is no unique
solution. In contrast, with the sparseness constraint in place, a
large fraction of the synaptic connections become zero (Figure~S2F),
and the optimal basis responses are uniformly spaced and almost
perfectly monotonic (Figure~S2C). These effects increase with the
strength of the penalty term (Figure~S2D and S2G).

To quantify the monotonicity of the resulting basis responses, the
derivative of each tuning curve was computed numerically, and a
monotonicity index for each curve was calculated from it.  Defining
$d_{ik} \eq \avg{r_{ik+1}} - \avg{r_{ik}}$,
the monotonicity index for cell $i$ is
\b
   MI_i = \frac{ \sum_k d_{ik} }{ \sum_l \left| d_{il} \right| } \, .
   \label{monoI}
\e
This number goes from $-1$ for a monotonically decreasing curve to
$+1$ for a monotonically increasing curve, with values near 0
indicating about equal numbers of increasing and decreasing steps.
Figure S2H--J shows how the distribution of monotonicity indices
changes as the strength of the sparseness constraint is increased. To
construct the histograms, optimal sets of ten tuning curves were
obtained six times with different initial conditions, producing
slightly different sets.  The 60 indices were then pooled to produce
the shown distributions.  The values are clustered near 0 when the
basis tuning curves oscillate and reach $\pm 1$ when they are
monotonic. 

The results in Figure~S2 show that the sparse-connectivity constraint
is effective at disambiguating the shapes of the optimal tuning curves
and may lead to monotonic profiles. Importantly, however, monotonic
curves are produced only when the target downstream functions are
themselves monotonic or have monotonic components. This is shown in
Figure~S3. When the desired downstream responses are oscillatory and
lack any directional bias, the resulting basis functions are
themselves oscillatory, and their monotonicity indices are clustered
around zero (Figure.~S3A and S3B). Most interestingly, for the
downstream functions used in Figure~3, which model hypothetical
reactions to binocular disparity signals, the result is again an
intermediate representation where the optimal tuning curves have
various degrees of monotonicity and thus a widely-spread distribution
of monotonicity indices (Figure~S3E).

\begin{figure*}[t!]
\centerline{\epsfig{figure=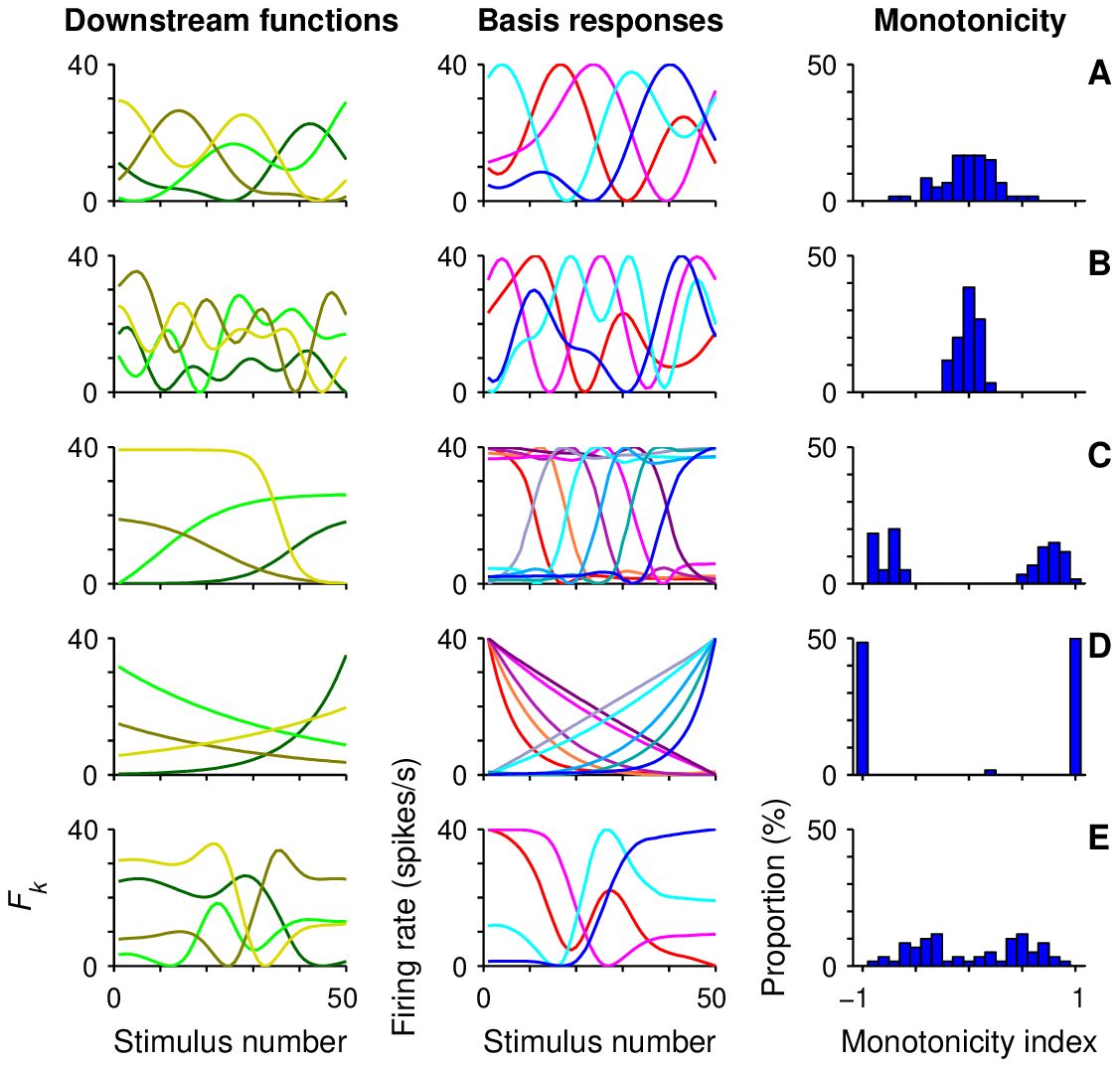,width=3.5in,clip} \hspace{0.10in}
\parbox[b]{2.40in}{\small
{\sf \textbf{Figure S3.}}
Optimal tuning curves for five classes of motor functions.
Results are as in Figure~S2, with a penalty strength of 0.002.  Each
row shows results for one class of downstream functions. Each panel in
the left column shows four representative examples of functions of a
given type; the middle column shows the resulting optimal basis
responses; the right column shows their distributions of monotonicity
indices.
\textsf{(A)} Low-frequency oscillating functions. For clarity, only
four of ten tuning curves are shown.
\textsf{(B)} As in \textsf{(A)}, but for high-frequency oscillating
functions.
\textsf{(C)} Increasing and decreasing sigmoidal curves (as in
Figure~S2C).
\textsf{(D)} Exponential functions.
\textsf{(E)} Motor functions that oscillate but have an increasing
or decreasing trend (as in Figure~3). \vspace{2pt} }}
\end{figure*}

\subsec{Influence of Stimulus Statistics}

One particular question that was thoroughly investigated using this
alternative set of constraints was whether the statistics of the
stimuli could affect the tendency of the basis neurons to develop
monotonic responses. The left-most column in Figure~S4 shows four
probability profiles that were tested.  The flat one at the top is the
standard, where all stimuli are equally probable ($s_k \eq 1/M$). In
the other three, some stimuli are a lot more frequent than others.
Sets of ten optimal basis responses were obtained with each profile
when the goal was to approximate either oscillatory or monotonic
downstream functions (Figure~S4A and S4B).  Figure S4 shows results
obtained using a high level of noise, which tended to enhance the
effects of the probabilities, as observed earlier with the parametric
approach (Figure~S2C and S2D). As can be seen by comparing the
resulting tuning curves in different rows, the basis responses deemed
optimal did adapt according to the probabilities.  In particular, with
non-monotonic downstream responses the variations in tuning curve
amplitude were a lot smaller in the regions where stimuli had low
probabilities. And with monotonic downstream responses it was the
spread of the tuning curves that varied appreciably. However, in both
cases the degree of monotonicity was barely affected by the stimulus
probabilities. The two columns under Figure~S4A show that when the
target functions are non-monotonic the optimal tuning curves should
also be non-monotonic, regardless of the probabilities $s_k$.
Similarly, the two columns under Figure~S4B indicate that when the
target functions are monotonic the optimal tuning curves should also
be predominantly increasing or decreasing, and the probabilities $s_k$
barely make a difference in this regard.  Therefore, it seems that
monotonic tuning curves cannot be generated simply by manipulating the
stimulus statistics. And furthermore, if monotonic curves are optimal
because of downstream requirements, it seems that the statistics of
the stimuli cannot override this trend.
\begin{figure*}[t!]
\centerline{\epsfig{figure=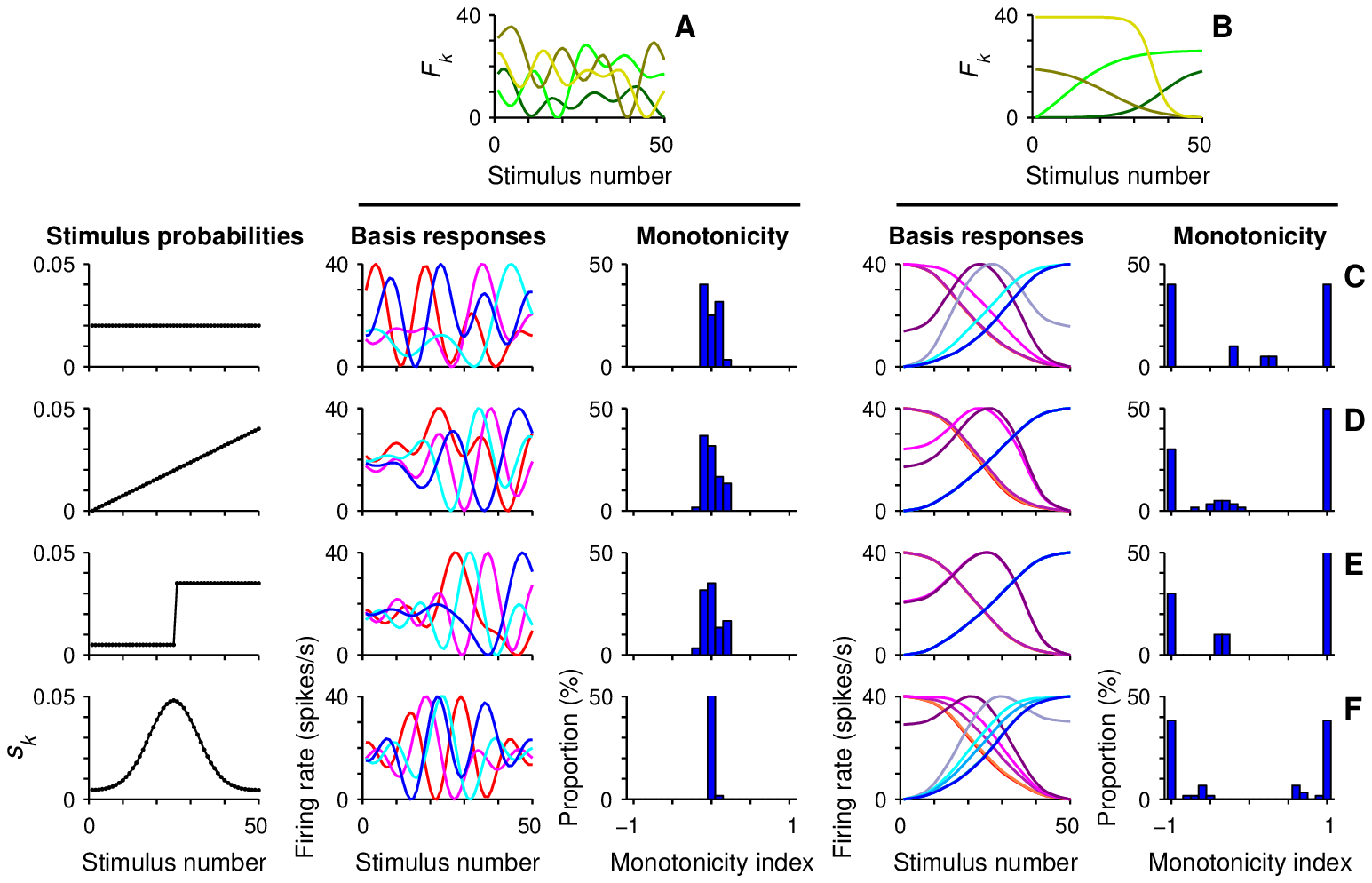,width=5.3in,clip}}
\vspace{0.1in}
\parbox[t]{\textwidth}{\small
{\sf \textbf{Figure S4.}}
Effect of stimulus statistics.
Sets of ten optimal basis responses were obtained for oscillatory
\textsf{(A)} and monotonic \textsf{(B)} downstream functions using
4 stimulus probability profiles (left-most column). A profile
corresponds to the probability $s_k$ as a function of stimulus $k$.
Each condition was repeated 6 times to obtain 60 monotonicity indices.
Optimal sensory responses and histograms of monotonicity indices are
shown for each condition, as indicated in the respective columns.
\textsf{(C)} Uniform profile. All stimuli were equally frequent.
\textsf{(D)} Linear profile. Stimulus probability increased steadily.
\textsf{(E)} Step profile. Stimulus probability changed abruptly at
stimulus 25.
\textsf{(F)} Gaussian profile. Stimuli in the middle of the range were
much more probable than at the edges.  All results are as in Figure~S3
(penalty strength of 0.002), except that high noise was used 
($\alpha \eq 0.5$). Monotonicity indices were highly insensitive to
stimulus probabilities.}
\end{figure*}

\sec{Match between Analytic and Numerical Results}

When a specific correlation matrix \FI\ is chosen,
Equations~\ref{EBerr2} and \ref{errSVD1} make a prediction about how
the mean square error should decrease as a function of the number of
tuning curves (basis functions) used for approximating the
corresponding downstream responses. Verifying that the numerical
results match these expressions is important, first, to check that the
minimization routines used to find the optimal basis functions are
working properly, and second, to investigate whether few basis
functions may indeed be sufficient to approximate accurately a large
number of downstream responses, as implied by the analysis.  

To address this, a special set of downstream responses was generated
which, by construction, could be approximated exactly by the family of
parameterized curves --- single-peak or monotonic --- used in
Figure~2.  This was done in three steps. (1) An arbitrary set of six
peaked and monotonic curves was generated using the regular
parameterization of Equation~\ref{tunc}; these six curves are shown in
Figure~S5A\@. 
\begin{figure*}[t!]
\centerline{\epsfig{figure=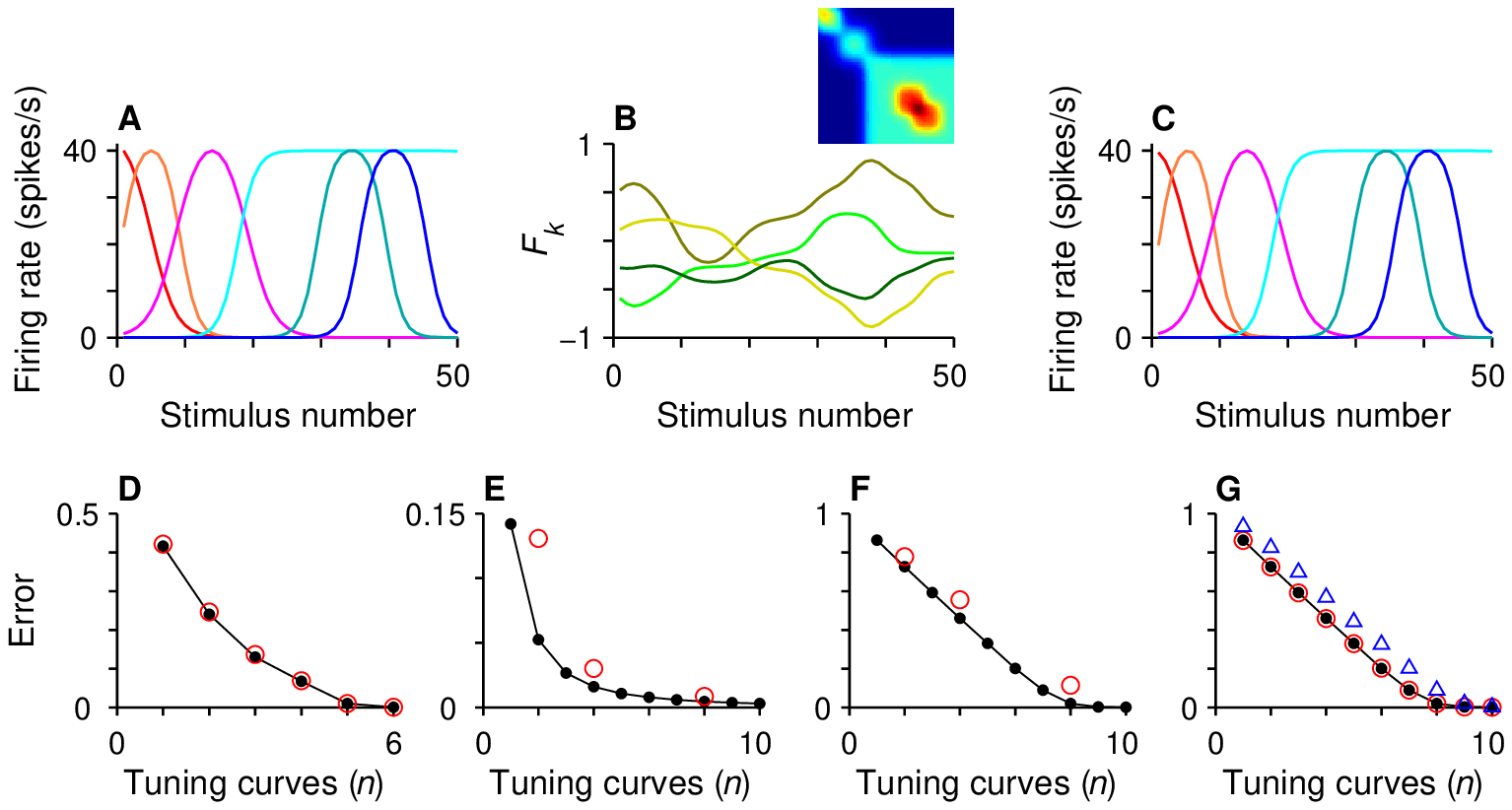,width=4.4in,clip}}
\vspace{0.1in}
\parbox[t]{\textwidth}{\small
{\sf \textbf{Figure S5.}}
Mean squared error in the approximation versus theoretical
minimum.
\textsf{(A)} Set of functions used to construct the downstream
responses shown in \textsf{(B)}, which served as controls. Each of
5000 downstream curves was generated as a linear combination of the
six responses shown in \textsf{(A)}, with random coefficients. The
resulting correlation matrix \FI\ is in the inset.
\textsf{(C)} Set of six optimal tuning curves recovered by the
minimization routine. The curves shown minimized \EB\ given the
control matrix \FI\ in panel \textsf{(B)} and zero noise.
\textsf{(D)} Mean squared error \EB\ as a function of the number of
optimal tuning curves, $n$, when the \FI\ matrix in \textsf{(B)}
was used. Black dots are the minimum errors calculated from
Equation~\ref{EBerr2}. Red circles are the \EB\ values obtained (from
Equation~\ref{EBerr}) with the indicated number of basis tuning
curves.
\textsf{(E)} As in \textsf{(D)}, but when the \FI\ matrix for
sigmoidal functions shown in Figure~2C was used. Red circles correspond
to the sets of 2, 4 and 8 curves shown in Figure~2G\@.
\textsf{(F)} As in \textsf{(E)}, but when the \FI\ matrix for
oscillatory functions shown in Figure~2A was used. Red circles
correspond to the sets of 2, 4 and 8 curves shown in Figure~2E\@.
\textsf{(G)} Optimal tuning curves for the oscillatory functions shown
in Figure~2A were computed using the alternative method with no
restriction on the synaptic weights (sparseness not enforced). Red
circles are for optimal tuning curves only constrained to have a
maximum of 1. Blue triangles are for curves constrained to have a
maximum of 1 and a minimum of 0. Black dots are the same as in
\textsf{(F)}. No noise was used in \textsf{(D)} and \textsf{(G)}.}
\end{figure*}
(2) A set of control downstream functions were constructed by adding
the six curves in random proportions. Thus, each downstream response
was equal to a linear combination of the six curves with coefficients
drawn randomly from $[-1,1]$. Four examples of such downstream
functions are shown in Figure~S5B, along with the resulting \FI\
correlation matrix. (3) This control \FI\ matrix was input to the
minimization routine, which searched for the sets of 1, 2,$\ldots$, 6
optimal tuning curves that minimized \EB. Note that the minimization
routine had no information about the downstream functions other than
\FI. The set of six optimal tuning curves ($n\eq6$) found by the
routine is shown in Figure~S5C\@.  They are almost identical to the
original set used to construct the control \FI. The mean square error
as a function of the number of optimal tuning curves used in the
approximation ($n$) is shown in Figure~S5D\@. The actual error values
(red circles) are superimposed on the minimum values (black dots)
expected from Equation~\ref{EBerr2}. The numbers are very close,
showing that the minimization routine indeed found the best possible
basis functions.  They are not exactly equal, particularly for the
first three points, because the shapes of, say, the best two basis
functions ($n\eq2$) cannot be perfectly fit by the parameterization of
Equation~\ref{tunc}. The error, however, is virtually zero for
$n\eq6$, where by construction the parameterized curves can indeed
reproduce the best basis functions exactly.

What happens with other correlation matrices \FI? An example is shown
in Figure~S5E, which plots the mean square error (red circles)
obtained in approximating the sigmoidal motor responses of Figure~2C
with the sets of 2, 4 and 8 tuning curves shown in Figure~2G\@. Black
dots are again the minimum values obtained from Equation~\ref{EBerr2},
except that now the eigenvalues of the \FI\ matrix in Figure~2C were
used. Note that the errors are generally small; just two basis
functions ($n\eq2$) capture more than 85\% of the variance of the
downstream responses.  This is because in this case all the motor
responses are rather similar to each other. In contrast, Figure~S5F
shows an analogous plot that was generated using the \FI\ matrix in
Figure~2A, which resulted from oscillatory motor responses. In this
case, both the minimum expected error (black dots) and the actual
approximation error obtained with 2, 4 or 8 tuning curves (red
circles) start much higher.  The difference between them, however, is
again small.

In Figure~S5E and S5F the red and black dots differ for two reasons,
first, because the noise was not zero, and any noise increases the
approximation error (Equation~\ref{errSVD1}), and second, because the
parameterized curves could not match exactly the shapes of the optimal
tuning curves. In all the examples studied, the error due to this
mismatch in shape was equivalent to about one tuning curve. Thus, the
error obtained with eight tuning curves was approximately equal to the
minimum error calculated with seven eigenvalues, and so on. This is
best illustrated in Figure~S5G, which was constructed as follows. The
black dots are the minimum errors for approximating the oscillatory
motor responses of Figure~2A; they are the same as in Figure~S5F\@.
The other data points are the errors obtained with $n$ tuning curves
found using the alternative method mentioned in the previous section.
That is, gradient descent was used to find sets of $n$ tuning curves
and connection weights that minimized \EL. Here, no noise and no
restrictions whatsoever were placed on the synaptic weights. The red
circles were obtained when the only constraint on the tuning curves
was that their maximum had to be equal to 1. This did not limit the
range of tuning curve values (which could be negative), nor the shape
of the profiles, so the approximation errors were virtually equal to
the theoretical minima. On the other hand, the blue triangles were
obtained with the same method, but when the tuning curves were
constrained to have a minimum of 0 and a maximum of 1.  This limited
their range of values and caused a small additional error. However,
with more than 4 or 5 tuning curves, the approximation error with $n$
tuning curves was almost exactly equal to the minimum (theoretical)
error with $n-1$ curves.

In conclusion, the families of downstream responses studied here can
indeed be described by a few eigenvalues and eigenvectors. As a
consequence, few tuning curves may be enough to approximate a large
variety of motor responses even when the range of tuning curve shapes
is limited.

\sec{Note on the Number of Motor Responses}

The \FI\  matrices shown here were produced by averaging 5000
downstream functions. Such large number was used so that, for a given
class of functions, the corresponding \FI\ matrices would change
little from run to run. With this number, two matrices generated with
different groups of 5000 functions from the same class were virtually
indistinguishable, and so were their eigenvalues. This was convenient
for eliminating a potential source of variability across runs, and
this in turn was important to ensure that the results were repeatable
and that the solution was not a local minimum of \EB.  However,
adequate (smooth) \FI\ matrices could also be obtained with 200
downstream functions or less, depending on their type.

The condition that there should be more motor neurons than sensory
neurons ($n\!<\!N$) is crucial for the present framework.  It is a
reasonable assumption if one considers that a given sensory stimulus,
say, a phone ringing, may trigger a variety of motor actions, such as
answering the phone, ignoring it, listening to the answering machine
to find out what the caller wants, etc. So, combined with the
different contexts in which a stimulus may appear, the number of
associated motor responses may be very large.

\end{document}